\documentclass[11pt]{article} 
\usepackage[english]{babel}
\usepackage{amsmath}
\usepackage{amssymb}
\usepackage{graphicx}
\usepackage{color}
\usepackage[utf8]{inputenc}
\usepackage{amsfonts}
\usepackage{graphicx} 
\usepackage{amsthm}
\usepackage{float}
\usepackage{fullpage}
\usepackage[colorlinks=true,linkcolor=red,citecolor=blue]{hyperref}
\usepackage{hyperref}
\usepackage{bm}
\usepackage[auth-sc-lg,affil-sl]{authblk}
\begin{document}
\newtheorem{defi}{Definition} \newcommand{\comment}[1] {\par {\bfseries \color{blue} #1 \par}}
\newcommand{\class}{(c)} \newcommand{\M}[1]{\mathbb{M}_{#1}}
\renewcommand{\vec}[1]{\underline{#1}}
\newcommand{\mat}[1]{\underline{\underline{#1}}}
\newcommand{\Ab}[1]{A_{{\rm b}#1}}
\newcommand{\As}[1]{A_{{\rm a}#1}}
\newcommand{\pb}{p_{\rm b}}
\newcommand{\mub}{\mu_{\rm b}}
\newcommand{\mua}{\mu_{\rm a}} \newcommand{\score}{\mathcal{S}}
\newcommand{\offerb}{\mathcal{O}_{\rm b}}
\newcommand{\offera}{\mathcal{O}_{\rm a}}
\newcommand{\proba}{\mathbb{P}} \newcommand{\sigmaa}{\sigma_{\rm a}}
\newcommand{\sigmab}{\sigma_{\rm b}}
\newcommand{\validorder}[2]{Q_{{\cal{#1}} #2}}
\newcommand{\acceptedorder}[2]{T_{{\cal{#1}} #2}}
\newcommand{\Nacc}[2]{\bar{N}_{\mathcal{#1} #2}}
\newcommand{\distn}[1]{\phi_{#1}} \newcommand{\A}{\underline{A}}
\newcommand{\dotA}{\underline{\dot{A}}} \newcommand{\Afbfs}[1]{A_{#1}}
\newcommand{\frate}{r} \newcommand{\payoff}{\mathcal{S}}
\newcommand{\fictitious}{\alpha} \newcommand{\noise}{\vec{\xi}}
\newcommand{\drift}{\vec{\mu}} \newcommand{\stdev}{\mat{\Sigma}}
\newcommand{\mcond}{\vec{m}_c} \newcommand{\ldev}{\sqrt{{r}}}
\newcommand{\logeq}{\underset{r\to0}{\asymp}}
\newcommand{\pricesetting}{\theta}
\newcommand{\avbid}{\langle\textrm{b}\rangle}
\newcommand{\avask}{\langle\textrm{a}\rangle}
\newcommand{\sigmoid}{\sigma_\beta} \newcommand{\Nb}[1]{N_{{\rm
      b},#1}} \newcommand{\Na}[1]{N_{{\rm a},#1}}
\newcommand{\pieq}{\pi_{\text{eq}}}
\newcommand{\Qam}{Q_{\mathrm{a},m}}
\newcommand{\Qbm}{Q_{\mathrm{b},m}}
\newcommand{\Tbm}{T_{\mathrm{b},m}}
\newcommand{\Tam}{T_{\mathrm{a},m}}
\newcommand{\Nta}[1]{\bar{N}_{\mathrm{a},#1}}
\newcommand{\Ntb}[1]{\bar{N}_{\mathrm{b},#1}}
\newcommand{\kernel}{\mathbb{K}} \newcommand{\strat}{a}
\newcommand{\gpayoff}[2]{\mathcal{P}^{(#1)} (#2)}
\newcommand{\player}{p}
\newcommand{\etal}{\textit{et al}.}
\newcommand{\ie}{\textit{i}.\textit{e}.}
\newcommand{\eg}{\textit{e}.\textit{g}.}
 \title{Dynamical selection of Nash equilibria
  using Experience Weighted Attraction Learning: emergence of heterogeneous mixed equilibria}
\author[1]{Robin Nicole \thanks{robin.nicole@kcl.ac.uk}}

\affil[1]{Department of Mathematics, King's College London, Strand, London, WC2R 2LS, United Kingdom}
\author[1]{Peter Sollich \thanks{peter.sollich@kcl.ac.uk}}
\maketitle
\begin{abstract}
  We study the distribution of strategies in a large game that models how agents choose among different 
  double auction markets. We classify the possible mean field Nash equilibria, which include potentially segregated states where an agent population can split into subpopulations adopting different strategies. As the game is aggregative, the actual equilibrium strategy distributions remain undetermined, however. We therefore compare with the results of Experience-Weighted Attraction (EWA)
learning, which at long times leads to Nash equilibria in the appropriate limits of large intensity of choice, low noise (long agent memory) and perfect imputation of missing scores (fictitious play). The learning dynamics breaks the indeterminacy of the Nash equilibria. Non-trivially, depending on how the relevant limits are taken, more than one type of equilibrium can be selected. These include the standard homogeneous mixed and heterogeneous pure states, but also \emph{heterogeneous mixed} states where different agents play different strategies that are not all pure. The analysis of the EWA learning involves Fokker-Planck modeling combined with large deviation methods. The theoretical results are confirmed by multi-agent simulations.
\end{abstract}

\section{Introduction}
Agent based models describe the dynamics of co-learning and interacting individuals
and can be applied in many fields including sociology -- with the
Schelling model of
segregation~\cite{doi:10.1080/0022250X.1971.9989794} a famous example -- and economics,
where the individuals are economic agents.  In recent decades,
there has been growing interest in the application of agent based
models to the study of financial markets; for extensive reviews of such applications we refer
to~\cite{doi:10.1080/14697688.2010.539249,samanidou2007agent}.
Among
existing models of double auction markets, one can
cite the work of Iori et al.~\cite{doi:10.1088/1469-7688/2/5/303} and
the CAT game~\cite{cai2009overview}. The latter is a market
design tournament in which participants were asked to supply automated
markets that would perform as well as possible in an economic system
populated with automated traders. Spontaneous emergence
of preferences for different markets emerged within the population of traders. Unfortunately, the complexity of the CAT game tournament
made it impossible to study this so-called segregation phenomenon by analytical methods, emphasizing the
need for a simpler model to understand the phenomenon of
segregation. Alori\'c et al.\ designed such a 
minimal version of the CAT game, where traders learn to choose among \emph{two} double auction markets~\cite{aloric2015emergence}. Also there segregation was observed, as the outcome of the learning dynamics. Whether this result has an interpretation as a game theoretical equilibrium was not addressed, however.
This will be one of the two main questions of this paper: we ask to
what extent segregation shows up in  the \emph{Nash equilibria} of the game corresponding to the model of Alori\'c
\etal{}. One of the properties of this game is that the payoff
agents earn by trading at the different markets depends only
on the ratio of the number of buyers and sellers at this market. The game therefore belongs to the class of aggregative games, where payoffs depend on a finite number of macroscopic quantities, called aggregates.

Bearing in mind the above broader context, we consider in this paper the double auction game of~\cite{aloric2015emergence} as a
paradigmatic example of an aggregative game with an infinitely large
number of players. While it is known that finding Nash equilibria
in games with a large but finite number of players is computationally hard
\cite{ne_complexity1},
taking the number of players to infinity can lead to drastic
simplifications that make the problem analytically tractable. This is because the limit eliminates some features such
as the market impact of the action of a single
player~\cite{ellison2003knife}. For aggregative games the limit also has convenient mathematical properties: Nash equilibria of infinite games can be characterized as the large size limit of equilibria in games with a finite number of players~\cite{carmona2004nash}. An introduction to games with a large
number of players would not be complete without mentioning mean field
game theory~\cite{Lasry2007,cardaliaguet2010notes},
which studies stochastic differential games with an infinite
number of players. The underlying formalism here is rather different from the one we use in the rest of this article, however.

Nash equilibria of aggregative game are characterized by the values of the aggregates on which the payoff of any given action depends. To each of these there generally correspond infinitely many different distributions of strategies among the
players.
In this paper, the second question we therefore ask is whether and how this degeneracy in the strategy distribution is resolved by the learning dynamics of the corresponding agent based model. This issue of how a 
Nash equilibrium is selected dynamically
has
been studied
theoretically for games of small
size~\cite{fudenberg1998theory} and using numerical simulation for larger games~\cite{kash2011multiagent,friedman1997learning,
cesa2006prediction},
providing results on the speed of convergence and efficiency of certain types of
learning dynamics. While these previous studies focused on the value of
macroscopic quantities such as the ratio of number of buyers to number of sellers once the learning dynamics has converged, we are interested in going further and
investigating the distribution of strategies, which is crucial in order to establish whether there is segregation or not. The specific learning rule we
study is  Experience Weighted Attraction (EWA) learning, which is well
known to reproduce quite accurately the behaviour of human subjects learning to play repeated normal form games~\cite{camerer1999experience}.
Strategies are encoded by so-called preferences in EWA learning, and the comparison of the \emph{preference distributions} that result from EWA learning dynamics with the properties of the underlying Nash equilibria is one of our main contributions; this is a novel approach that has not to our knowledge been pursued in the existing literature.

Methodically, we argue that in the game we analyse, correspondence with Nash equilibria requires a long memory limit. 
The EWA dynamics of the agents is then described by a Fokker-Planck equation, and it is the steady states of this that we study. We deploy large deviation methods to detect segregation, where agents
split into sub-populations that each play a different
strategy. We combine this approach with numerical simulations in order to
shed light on the
several, qualitatively different, types of preference distribution that can emerge in the
steady state of the learning dynamics. These include the two scenarios that are conventionally considered:
homogeneous mixed equilibria, where all agents play the same mixed strategy, and heterogeneous pure equilibria, where different agents play different pure strategies~\cite{cabral1998,Schmeidler1973,rath1992direct}. Surprisingly, however, we also find heterogeneous mixed solutions, where the agents play different strategies and these strategies themselves include mixed strategies.

This paper is organized as
follow. In Sec.~\ref{sec:model} we summarize the minimal model of traders choosing between double
auction markets to be studied in the rest of this article, as well as the
EWA learning dynamics. In Sec.~\ref{sec:mf} we
study the Nash equilibria of the aggregative game corresponding to this model, in the limit of a large number of players.
In
Sec.~\ref{sec:learn} we present a study of the steady states of the
learning dynamics in the model of Sec.~\ref{sec:model} and argue that in the limit
of
\emph{fictitious play}, \emph{best
  response dynamics} and \emph{large memory}, these steady states are Nash
equilibria. We show that depending on how these multiple limits are approached, the dynamics selects several distinct Nash equilibria, including ones of heterogeneous mixed type.
In Sec.~\ref{sec:method} we present separately  the large
deviation methods that we use in our study of the steady states of EWA
learning in the large memory limit. Sec.~\ref{sec:conclusion} summarizes our results and lays out some avenues for future research. Technical details are relegated to the appendixes.

\section{Model: Choosing between Double Auction Markets}
\label{sec:model}
In this section, we summarize the model of double auction markets of
Alori\'c \etal~\cite{aloric2015emergence}. In this model, a population of co-evolving traders competes to trade by choosing between two double auction markets. This can lead to segregation, where agents spontaneously split into groups with different preferences for the two markets.
The model contains three ingredients: (i) the market
mechanism by which the double auction markets process orders to buy and sell, (ii) the way traders set their order prices (this is assumed fixed and not affected by learning) and calculate their payoff,
and (iii) the learning procedure that traders use to learn
their trading strategy, i.e.\ their preference for each market. We describe these three ingredients in turn.

\subsubsection*{Market mechanism}
The model assumes that each market processes orders in discrete trading rounds rather than continuously. In each round each trader places at one of the markets an order to buy or sell one unit of the underlying good.
An order is denoted $(\tau,p)$ where $\tau \in \{\mathrm{a},\mathrm{b}\}$ designates the type
of order, with $\mathrm{a}$ an order to sell
(also known as an ask) and $\mathrm{b}$ an order to  buy (a bid); $p$ is the price at which the trader proposes to
buy or sell. For example $(\mathrm{b},20)$ is an order to buy one
unit of good at a price of $20$. Once all the traders have sent their
orders
(see Dynamics of traders), the clearing process begins. The trading price is set by each market using the formula
\begin{equation}
  \pi_m = (1 - \theta_m) \avbid + \theta_m \avask
  \label{eq:tp}
\end{equation}
where $\avbid$, $\avask$ are the average prices of bids and asks received by
the market. All the orders on the wrong side of the trading price
(\ie{} an order to buy lower than the trading price or an order to
sell higher than the trading price) are rejected. The remaining \emph{valid orders}
are \emph{executed} at the trading price by randomly forming pairs of one buyer and one seller until no more pairs can be formed. As the
number of valid bids and asks will differ in general, some traders will remain unmatched; they are unable to trade and their orders are not executed.

\subsubsection*{Order pricing and payoff calculation} As explained above, it is assumed that
traders \emph{always} send an order to buy or sell \emph{exactly} one
unit of good to only one single market. This is done to keep the model as simple as possible. Following the work
of Gode and Sunders~\cite{gode1993allocative}, traders set the price
of their orders with \emph{zero intelligence}: the
price of each order to buy (resp.~sell) sent by each trader is an independent
Gaussian random variable with mean $\mub$ (resp.~$\mua$) and standard
deviation $\sigmab = \sigmaa = 1$. While this assumption may 
appear drastic at first sight, Gode and Sunders found that traders
sending orders to double auction markets with zero intelligence was a
good substitute for individual
rationality~\cite{gode1993allocative}.
The model also assumes that each agent chooses randomly whether to buy or sell, with a fixed probability $\pb$ that can be different for different agents.

At the end of a trading round,
each trader receives as feedback from the market to which they sent their
order whether it was executed and if so at which price.
From this each trader computes the score of his order $\mathcal{S}$ as either
zero, if the order was not executed, or otherwise as the profit of the
order, which in the model is defined as the absolute value of the difference between order price and trading price. This payoff is random and is affected  by:
(i) the submitted order price, (ii) the trading price, and (iii) whether the order is executed, which in turn depends on 
the ratio of number of buyers and sellers in the market where the offer was sent.
(We discuss in Sec.~\ref{sec:mf} how the average payoff over these sources of randomness can be calculated in the limit of a large system.)

\subsubsection*{Dynamics of traders}

The remaining part of the behaviour of the traders that the model needs to prescribe is how they
learn their respective preferences for the two markets. The assumption is that agents use
experience-weighted attraction reinforcement learning (EWA)
\cite{camerer1999experience}. They have attractions $\Afbfs{m}$ to
each market $m \in \{1,2\}$, which they update after each trading
round
$n$ according to
\begin{equation}
  \Afbfs{m}(n+1) =\left\{
    \begin{array}{cc}
      (1 - \frate) \Afbfs{m}(n) + \frate \payoff(n) & \text{if the agent chose market $m$ in round $n$}\\
      (1 - \fictitious \frate) \Afbfs{m}(n) & \text{otherwise}                                                      
    \end{array}\right.  
\label{eq:EWA_dynamics}
\end{equation}
Here $\payoff(n)$ is the payoff for the order placed at time-step $n$,
$\alpha$ is a \emph{fictitious play parameter} which describes how fast
traders decrease the attraction to actions they do not play, and
$r$ is the inverse of the agents' memory, defined as the period of time over which they typically remember past payoffs. Based on those
attractions $\mathbf{A}=(A_1,A_2)$, traders then randomly choose a market for trade according to the inverse logit or ``softmax'' function $\sigma_\beta(\cdot)$,
\begin{equation}
  \label{eq:ptrade}
  \proba(\text{trade at market }1 \mid \mathbf{A} ) =\sigmoid(A_1 - A_2) = \frac{1}{1+\exp(-\beta (A_1 - A_2))}
\end{equation}
where $\beta$ is the intensity of choice that regulates how strongly the agents use the attractions to bias their preferences.
A possible extension of this setup, which we do not pursue here, is to allow the traders to learn also their preference for buying and selling, instead of keeping this fixed~\cite{aloric2015emergence}. 
In that case there would be four attractions to be learned, for buying and selling at each of the two markets.

We shall use ``EWA learning'' as a shorthand to designate the above dynamics where traders learn
at which market to trade -- note that because of this learning process the traders are somewhat more intelligent than the strictly zero-intelligence traders described
by Gode and Sunders~\cite{gode1993allocative}, who in our scenario would choose randomly also where to trade.

\begin{figure}[h!]
  \centering
  \includegraphics[scale = 0.6]{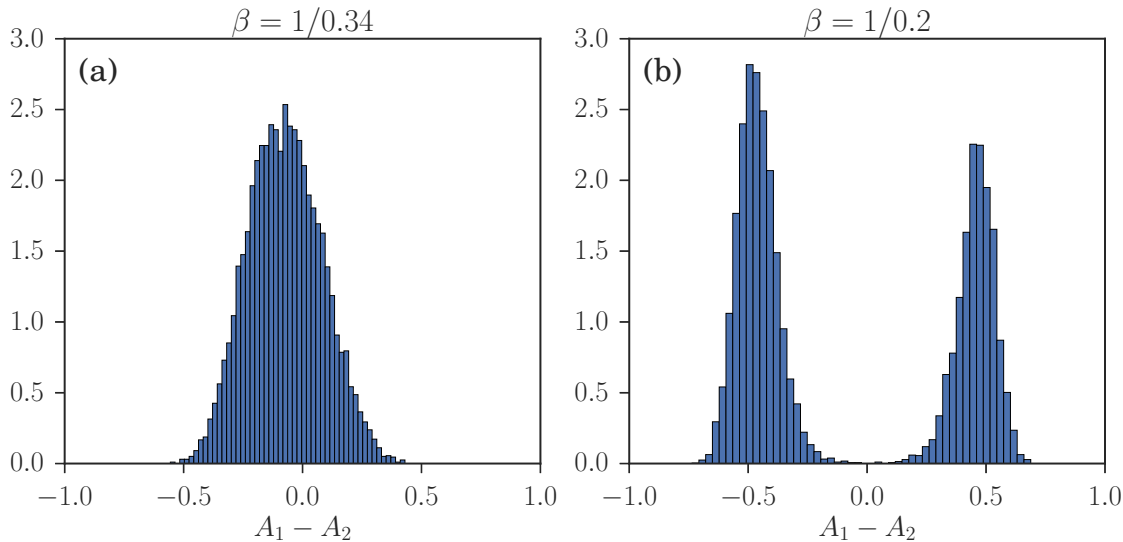}
  \caption{Results of a multi-agent simulation of the model of~\cite{aloric2015emergence} after $5\cdot 10^4$
rounds of trading among $2\cdot10^4$ agents.  Parameters for the two markets are
    $\theta_1 = 1- \theta_2 = 0.3$, buying preferences for the two classes of agents are $\pb^{(1)} = 1 - \pb^{(2)} = 0.2$, forgetting rate $r = 0.01$ and $\alpha=1$ (no fictitious play). 
Shown is the distribution of attraction differences $A_1-A_2$ across the first group of agents.  This is unimodal for intensity of choice $\beta$ below the
segregation threshold as in (a)
, but becomes bimodal for larger $\beta$: the system shows emergent segregation.
}
\label{fig:simuAloric}
\end{figure}

In the following we focus largely on a symmetric setup~\cite{aloric2015emergence}, explained in more detail in Sec.~\ref{sec:classific} below. There are two classes of agents in this scenario but their 
distributions of attractions are related by swapping $A_1$ and $A_2$ so it is enough to focus on one class.
Numerical simulation and theoretical analysis of EWA
learning, for $\alpha=1$, then show that when the intensity of choice $\beta$ is
above a threshold $\beta_c$ the distribution of the traders' attractions can become bi-modal~\cite{aloric2015emergence}: the model produces emergent segregation. By way of orientation, example simulation results for $\beta$ both below and above the segregation
threshold are shown in Fig.~\ref{fig:simuAloric}.
\subsubsection*{Incomplete versus complete information}
One possible cause of heterogeneity in agents' preferences that has been identified in previous studies is incomplete or imperfect information~\cite{MadhavanFragmentation}.
An obvious question is whether this explains the observation of segregation in the double auction market model described above.
Indeed, the agents in this model do have incomplete information
about the
markets they are trading in: they only receive the stochastic payoffs but do not have access to global information such as the number of buyers and sellers at each market, which they would need in order to estimate their average payoff.
As a consequence, traders face the
exploration/exploitation dilemma that is common in 
reinforcement learning~\cite{RLIntro}.
They need to \emph{explore} the
whole strategy space (both high and low payoff strategies) to have
accurate payoff estimates for their strategies, while at the same
time \emph{exploiting} the most profitable strategy by playing it frequently. In the model we consider the trade-off between exploration and exploitation is set by the intensity of choice $\beta$~\cite{galla2013complex}, with higher values favoring exploitation by making agents choose predominantly the market with the larger attraction.

To address the question of whether segregation is possible also with \emph{perfect} information, we develop in the next section an appropriate game theoretical version of the double auction model discussed above. Once we have determined the Nash equilibria of this game, we will
come back to a comparison with the steady state of the EWA learning dynamics, to see how this resolves an indeterminacy in the Nash equilibria.

\section{Mean field Nash equilibria}
\label{sec:mf}
We now rephrase the double auction market choice model of Sec.~\ref{sec:model} in game theoretical language. This will allow us to determine and classify its Nash equilibria in the mean field limit of an infinite number of players. Our aim will be to determine whether in this \emph{perfect information} context there are still signatures of the segregation phenomenon previously found for EWA learning with imperfect information.
In Sec.~\ref{sec:learn} we will then see that, in the appropriate limit, the steady states of the EWA learning dynamics are consistent with the Nash equilibria of the model described in this
section.

\subsection{Game theoretical framework}
\label{sec:formalism}

\subsubsection*{Setting} We consider a population of $N $ traders
called players (to be consistent with standard terminology in game
theory). Those players are divided into two classes $c \in \{1,2\}$, of the same size.  Each player has fixed
buy/sell preferences described by the probability to buy, $\pb^{(c)}$, which depends on
his/her class. Each trading round is a round of the game, where each player chooses one of two actions, viz.\ ``send an order to market one'' and ``send an order to market two''; we label these by
$m \in \{1,2\}$. A \emph{pure strategy} is one where a player always chooses the same action. A \emph{mixed strategy} is one where the player chooses action $m=1$ with probability $p \in [0,1]$ and $m=2$ otherwise.
This formalism can be linked to EWA learning as described in
Sec.~\ref{sec:model}: there the traders learn which mixed strategy to
play, mapping the learned attractions $(A_1,A_2)$ to the probability $p$ using the softmax function $\sigma_\beta(\cdot)$ defined in Eq.~\eqref{eq:ptrade}.
\subsubsection*{Average payoff in a large game}

To determine the Nash equilibria, we need to determine the average payoff of a player for a given strategy $p$, given the (fixed) strategies of all other players. While this calculation would be complicated for finite $N$, it simplifies in the limit $N\to\infty$ that we consider from now on. Firstly, the trading price at each market becomes non-fluctuating as the average value of bids and asks submitted becomes equal respectively to $\mub$ and $\mua$, up to fluctuations that vanish as ${\cal O}(1/\sqrt{N})$. 

Secondly, the ratio of the number of buyers and sellers at each market $m$, which we denote $f_m$, also becomes non-fluctuating. We can calculate these ratios from the strategy distribution $\phi^{(c)}(p)$ within each class of players, where because of the large $N$-limit we can neglect the effect of the strategy chosen by of any single player to obtain
\begin{subequations}
  \begin{align}
    \label{eq:2}
    f_1(\phi^{(1)},\phi^{(2)}) &= \frac{\pb^{(1)} \bar{p}^{(1)} + \pb^{(2)} \bar{p}^{(2)}}
                                 {(1-\pb^{(1)}) \bar{p}^{(1)} + (1-\pb^{(2)}) \bar{p}^{(2)}}\\
    f_2(\phi^{(1)},\phi^{(2)})  &= \frac{\pb^{(1)}(1-\bar{p}^{(1)}) + \pb^{(2)} (1-\bar{p}^{(2)})}
                                  {(1-\pb^{(1)}) (1-\bar{p}^{(1)}) + (1-\pb^{(2)}) (1-\bar{p}^{(2)})}
  \end{align}
\end{subequations}
Here $\bar{p}^{(c)} = \int {\rm d} p\, \phi^{(c)}(p) p $ is the
average mixed strategy parameter $p$ in class $c$.
In the above formulas, $N\pb^{(1)} \bar{p}^{(1)}$ is the typical number of agents of class 1 choosing to buy and to send their buy order to market 1. The relative fluctuations of this number again vanish for $N\to\infty$. The other terms in the expressions for the $f_m$ have analogous interpretations, and the common factor of $N$ cancels.

Based on the above considerations, it becomes a simple matter to calculate the average payoff
$\mathcal{P}_{\tau,m}(f_m)$ of buying $(\tau = \mathrm{b})$ or selling
($\tau = \mathrm{a}$) in market $m$, depending on the market conditions as encoded by $f_m$ (see
Appendix~\ref{app:PayoffFormula}).  Our game is therefore \emph{aggregative}~\cite{AggregativesCorchon}: average payoffs are determined only by the \emph{aggregate} quantities $f_1$ and $f_2$ that can be calculated from the strategy distributions $\phi^{(c)}(p)$. Other games in this class include the Cournot
oligopoly; in statistical physics language the aggregates would be called order parameters. 

In our setup we need to average the payoff
$\mathcal{P}_{\tau,m}(f_m)$ further over the probability of buying or selling, giving for a player of class $c$ an average payoff for the action of ``going to market $m$'' of
\begin{equation}
  \mathcal{P}^{(c)}_m(f_m) = \pb^{(c)} \mathcal{P}_{\mathrm{b},m }(f_m) + (1-\pb^{(c)})\mathcal{P}_{\mathrm{a},m} (f_m) 
\end{equation}
Finally, for a player using a mixed strategy, the resulting payoff $\mathcal{P}^{(c)}(p , f_1,f_2)$ is an average
of the payoff at market 1 weighted by $p$ and the payoff at market $2$
weighted by $1-p$:
\begin{equation}
  \mathcal{P}^{(c)}(p, f_1,f_2) =  p \mathcal{P}^{(c)}_1( f_1) + (1-p) \mathcal{P}^{(c)}_2( f_2)
  \label{final_payoff_average}
\end{equation}
This quantity is the key input into the calculation of the Nash equilibria of our game.

\subsubsection*{Nash equilibria}

We choose to use the following definition of a Nash equilibrium for
our game in the limit of an infinite number of players~\cite{cardaliaguet2010notes}. This definition takes
advantage of the fact that we exploited in the payoff calculation, namely that for $N\to\infty$ the
aggregate quantities $f_1$ and $f_2$ remain constant if a single
player changes strategy; in other words, players do not have market impact and their payoff depends only on their own strategy and
the \emph{distribution} of the strategies in the population overall.

\begin{defi}{Nash equilibrium:}
  The strategy distributions $\phi^{(1)}$ and
  $\phi^{(2)}$ constitute a Nash equilibrium of the game if the two following conditions are verified: 
  \begin{subequations}
    \begin{align}
      \mathrm{Support}(\phi^{(1)}) \subseteq \mathrm{argmax}_{p}\left(
      \gpayoff{1}{p,f_1(\phi^{(1)},\phi^{(2)}),f_2(\phi^{(1)},\phi^{(2)}})\right)
      \label{eq:ne1}\\
      \mathrm{Support}(\phi^{(2)}) \subseteq \mathrm{argmax}_{p}\left(
      \gpayoff{2}{p,f_1(\phi^{(1)},\phi^{(2)}),f_2(\phi^{(1)},\phi^{(2)}})\right) \label{eq:ne2}
    \end{align}
  \end{subequations}
  Here the maximization of the payoff on the right hand side is performed over the variable $p$ at constant $\phi^{(c)}$; \ie{} each
single player maximizes their payoff with the aggregate quantities fixed.
\end{defi}
In words, the definition means that any strategy that has nonzero probability of being played by a player from class
$c$ (\ie{} in the support of $\phi^{(c)}$) must maximize the player's payoff. We will now apply this definition to determine the different
classes of Nash equilibria that exist in the double auction market choice game.

\subsection{Classification of Nash equilibria}
\label{sec:classific}

\subsubsection*{Equal payoff constraints}

We will classify Nash equilibria according to two
characteristics. If all agents in a class play the same strategy $p=\bar{p}^{(c)}$, the distribution $\phi^{(c)}(p)$ is a delta-distribution $\delta(p-\bar{p}^{(c)})$ and we call the equilibrium \emph{homogeneous} for that class, otherwise---when different players in the same class use different $p$--- we refer to the equilibrium as \emph{heterogeneous}. The second characteristic is the strategy type: if all agents in a class play the pure strategies $p=0$ or $p=1$ we call the equilibrium \emph{pure}, otherwise \emph{mixed}. Combining these two characteristics then divides equilibria for each class into four possible types.

To obtain a classification of the possible overall Nash equilibria, note that the function being maximized in
Eq.~(\ref{eq:ne1},\ref{eq:ne2}), viz.
$p \to
\gpayoff{c}{p,f_1(\phi^{(c)},\phi^{(2)}),f_2(\phi^{(1)},\phi^{(2)}})$
is \emph{linear} in $p$. As a
consequence, if it is not constant, it has a single maximum on
one of the boundaries of the interval $[0,1]$ where it is defined. 
A glance at~\eqref{final_payoff_average} shows that the payoff function
is constant if and only if $\phi^{(1)}$ and
$\phi^{(2)}$ are such that the payoffs at the two markets are equal:
\begin{equation}
  \mathcal{P}^{(c)}_1\left(f_1(\phi^{(1)},\phi^{(2)})\right) = \mathcal{P}^{(c)}_2\left(f_2(\phi^{(1)},\phi^{(2)})\right) \label{eq:eqpayoffcond}
\end{equation}
If (and only if) this \emph{equal payoff condition} is satisfied, the strategy distribution $\phi^{(c)}(p)$ can be nonzero for any $p\in [0,1]$. This can be interpreted by saying that, if in a class there are players that go the first and the second market, the only way for none of
them to have an incentive to move to another market is for the payoff at the two markets to be the same.

If the equal payoff condition is not met for a class, we have to have either
\begin{equation}
  \mathcal{P}^{(c)}_1\left(f_1(\phi^{(1)},\phi^{(2)})\right) > \mathcal{P}^{(c)}_2\left(f_2(\phi^{(1)},\phi^{(2)})\right), \qquad
\phi^{(c)}(p)=\delta(p-1), \qquad
\bar{p}^{(c)}=1
\end{equation}
or
\begin{equation}
  \mathcal{P}^{(c)}_1\left(f_1(\phi^{(1)},\phi^{(2)})\right) < \mathcal{P}^{(c)}_2\left(f_2(\phi^{(1)},\phi^{(2)})\right), \qquad
\phi^{(c)}(p)=\delta(p), \qquad
\bar{p}^{(c)}=0
\end{equation}
In both cases the strategy distribution is homogeneous pure, and the entire class of agents goes to the market with the higher payoff.

\subsubsection*{Types of Nash equilibria}

\begin{figure}[t!!]
  \centering
  \includegraphics[scale=0.5]{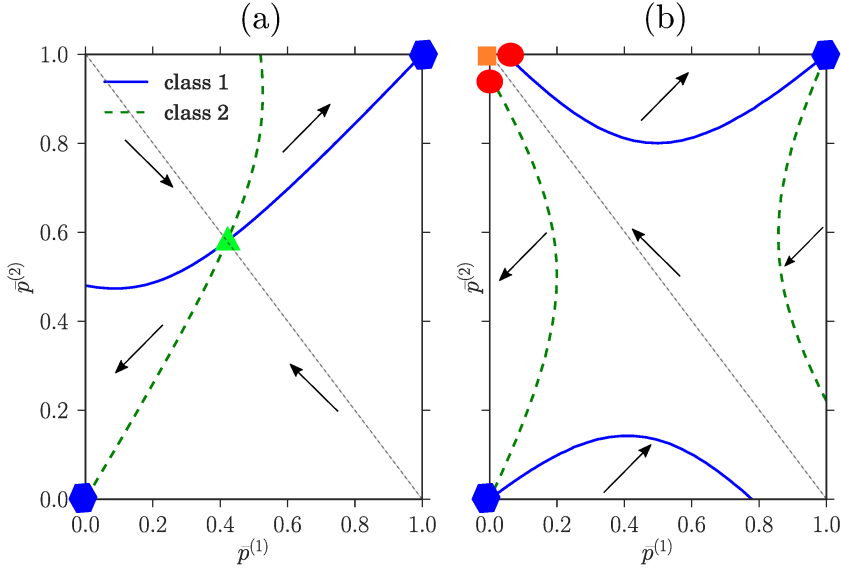}
  \caption{Values of $\bar{p}^{(1)}, \bar{p}^{(2)}$ for which the equal
    payoff constraints are verified for class $c=1$ (blue, solid) and class $c=2$
    (green, dashed). The arrows point to $(s^{(1)},s^{(2)})$ where
    $s^{(c)} \in \{0,1\}$ indicates the profit-maximizing strategy of traders from
    class $c$, in each distinct area of the plane.
In panel (a) where
    $\theta_1 = 1 - \theta_2 = 0.3$,
    $\pb^{(1)} = 1 - \pb^{(2)} = 0.2$, there exists a 
    heterogeneous 
    equilibrium (green triangle), located
    at the intersection of the two equal payoff curves.  In panel (b),
    $\theta_1 = 1 - \theta_2 = 0.2$,
    $\pb^{(1)} = 1 - \pb^{(2)} = 0.45$, and the equal payoff curves do not
    cross. There is then no potentially heterogeneous Nash
    equilibrium, but the direction of the arrows shows that a homogeneous
    pure equilibrium (orange square) with the two classes going to different
    markets exists. There are also two 
    partially heterogeneous Nash
    equilibria (red circles, see main text).
    In both (a)
    and (b) there exist homogeneous pure Nash equilibria
    where the whole population trades at the same market (blue hexagons). The dotted line indicates the location of the symmetric equilibria that we mostly focus on.
  }
  \label{fig:eqpayoff}
\end{figure}

We can now proceed to find the possible types of overall Nash equilibria for our game. Because $f_1$ and $f_2$ are fixed once $\bar{p}^{(1)}$ and
$\bar{p}^{(2)}$ are known, the equal payoff condition for each class defines a line of points in the $(\bar{p}^{(1)},\bar{p}^{(2)})$ plane. This line can consist of several distinct pieces as shown in the examples in 
Fig.~\ref{fig:eqpayoff}, where equal payoff lines are plotted for both
class $c=1$ (full lines) and $c=2$ (dashed lines).

\begin{figure}[t!]
  \centering
  \includegraphics[scale = 0.4]{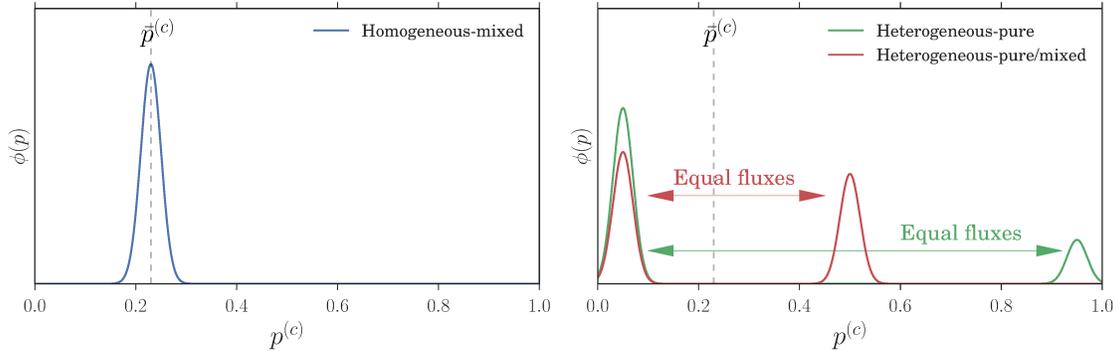}
  \caption{Three different types of strategy distribution $\phi(p)$ that 
all have the
    same mean $\bar{p}$ (dashed line): homogeneous mixed distribution
    (left panel), heterogeneous mixed (red curve, right panel)
    heterogeneous pure (green curve, right panel). Peaks in the distribution are shown broadened as they would be in EWA learning at finite decision strength $\beta$; as Nash equilibria they would become sharp (delta-distributions). The right panel illustrates that, when a strategy distribution has two distinct peaks, it can represent a steady state of the learning dynamics only when the fluxes of agents moving from one peak to the other balance in the two directions
(see Sec.~\ref{sec:method}).}
  \label{fig:diff_ne}
  \label{fig:1}
\end{figure}

The discussion above can now be summarized in graphical terms as follows: a point in the $(\bar{p}^{(1)},\bar{p}^{(2)})$-plane is a Nash equilibrium if for each class the point is either on the equal payoff line, or on the boundary (specified by $\bar{p}^{(c)}=1$ or $=0$) corresponding to the market where the class has the higher payoff. Combining these options for the two classes, the first and for our purposes most interesting type of Nash equilibrium that results is a point at an intersection of two equal payoff lines, away from the boundaries. We call such a point a \emph{potentially heterogeneous} Nash equilibrium. Here both $\bar{p}^{(1)}$ and $\bar{p}^{(2)}$ are strictly between 0 and 1. The strategy distributions can then be either
\begin{itemize}
\item homogeneous mixed, with $\phi^{(c)} = \delta(p-\bar{p}^{(c)})$, or
\item heterogeneous pure, with $\phi^{(c)} = (1-\bar{p}^{(c)})\delta(p)+
\bar{p}^{(c)}\delta(p-1)$, or
\item heterogeneous mixed otherwise.
\end{itemize}
These three different cases are illustrated schematically in Fig.~\ref{fig:diff_ne}.
The homogeneous mixed case can be viewed as the Nash equilibrium analogue of the unimodal distribution in the stochastic simulations shown in Fig.~\ref{fig:1}; in the heterogeneous mixed case the strategy distribution is arbitrary except for its fixed mean $\bar{p}^{(c)}$. The fact that the Nash equilibrium conditions here allow both homogeneous and heterogeneous strategy distributions motivates our use of the term ``potentially heterogeneous''. It also shows that 
one needs dynamical information to say more about the strategy distribution shapes, as explored in detail in Sec.~\ref{sec:learn}.

A second type of Nash equilibrium results when the equal payoff condition is obeyed for only one class while the other class is at a boundary. We then speak of a \emph{partially potentially heterogeneous} Nash equilibrium, because one class of players has a homogeneous pure strategy distribution while the other strategy distribution is of one of the three types listed in the bullet points above.

Finally, Nash equilibria unconstrained by either of the equal payoff
conditions must be in on of the four corners of the square
$(\bar{p}^{(1)},\bar{p}^{(2)})\in [0,1]^2$; we call them
\emph{homogeneous pure} equilibria as the strategy distributions for
both classes are then of this type. These equilibria can be further
subdivided depending on whether  both classes go to the same market or not. The former type always
exists as if one of the traders tries to trade in the empty market
s/he will earn a payoff of $0$ which is smaller than the payoff s/he
could earn in the non-empty market. In the latter type, each market is used only by traders of one class, who trade with each other there.

Plots in the $(\bar{p}^{(1)},\bar{p}^{(2)})$-plane as shown in 
Fig.~\ref{fig:eqpayoff} are a convenient graphical tool to assess the
existence of potentially heterogeneous, potentially partially heterogeneous and 
homogeneous pure Nash equilibria. Potentially heterogeneous equilibria are found directly as interior crossing points of the equal payoff curves for the two classes.
A partially heterogeneous Nash equilibrium corresponds
to a point (see Fig.~\ref{fig:eqpayoff}(b)) that is located at the
intersection of the equal payoff curve of class $1$ (resp.\ 2) and a horizontal 
(resp.\ vertical) boundary. This criterion identifies a list of (usually four) candidate equilibria. To have an actual equilibrium the payoffs of the markets for the homogeneous pure class need to have the correct order, e.g.\ 
for a candidate point located on the axis
$\bar{p}^{(2)}=1$, the payoff at market $1$ has to be higher for class 2 players than the payoff at market $2$. By drawing arrows indicating payoff ordering as explained 
in the caption of Fig.~\ref{fig:eqpayoff}, this can be summarized by saying that the arrows must point \emph{towards} the boundary that a candidate point for a potentially partially heterogeneous Nash equilibrium lies on. In Fig.~\ref{fig:eqpayoff}, this leaves two equilibria of this type as marked by the red circles.

Finally, for a heterogeneous pure Nash equilibrium where the two
classes of players choose different markets, the two candidate points
are the top left or bottom right corner. These are again Nash
equilibria provided they have the correct ordering of payoffs, which
requires that the arrows drawn in the figure point towards this
corner. In Fig.~\ref{fig:eqpayoff}(b) this is the case for the top left corner (orange square).

\begin{figure}[t!]
  \centering
  \includegraphics[scale = 0.8]{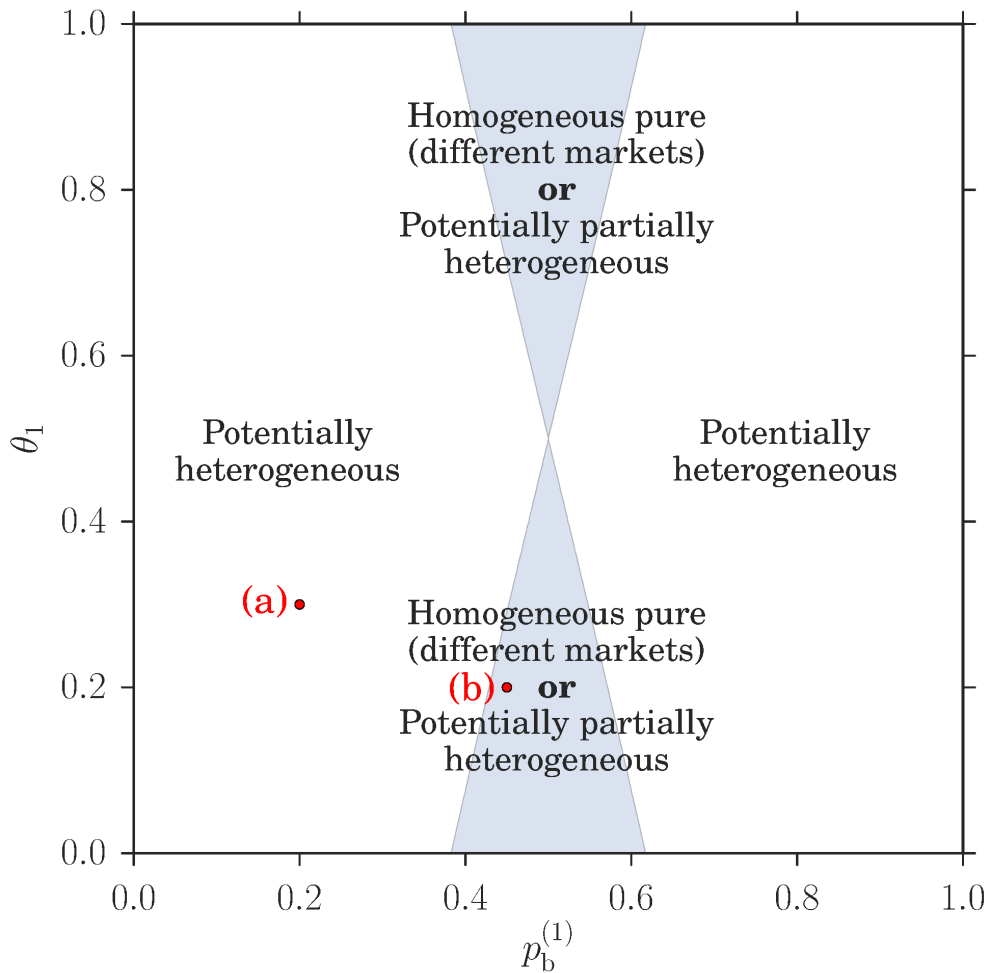}
  \caption{Phase diagram for existence of different types of
    Nash equilibria for a system with symmetric price setting
    parameters $\theta_1 = 1 - \theta_2$ 
    and buying preferences
    $\pb^{(1)} = 1 - \pb^{(2)} $. The types of equilibria in this plot are
    explained in Sec.~\ref{sec:classific} and a graphical method to
    check their existence is shown in Fig.~\ref{fig:eqpayoff}. The
    labels (a) and (b) correspond to the panels there.
Note that the two homogeneous
    pure Nash equilibria where both classes of player trade at the same
    market are not shown as they exist everywhere.
}
  \label{fig:phidiag}
\end{figure}

We can now look at how the existence of the different types of Nash equilibria depends on the system parameters, which are the market biases $\theta_m$ and the buying preferences $\pb^{(c)}$. We follow Ref.~\cite{aloric2015emergence} in focusing on a symmetric setup where the two markets have opposite biases in favour of buyers and sellers. As $\theta=0.5$ corresponds to the absence any bias, this means $\theta_1+\theta_2=1$. Similarly we assume that the players fall into two symmetric groups with respect to their buying preferences, with those in class 1 preferring to buy ($\pb^{(1)}<0.5$) and the others having the opposite preference $\pb^{(2)}=1-\pb^{(1)}$. With these choices, we can show 
in
Fig.~\ref{fig:phidiag} the regions where the different types of Nash equilibria exist as a function of $\pb^{(1)}$ and $\theta_1$. It turns out that the two examples shown in Fig.~\ref{fig:eqpayoff} cover the two generic cases: in addition to homogeneous pure Nash equilibria where both classes go to the same market, which always exist, one has either a potentially heterogeneous Nash equilibrium as in Fig.~\ref{fig:eqpayoff}(a), or a homogeneous pure equilibrium with the two classes at different markets and two potentially partially heterogeneous equilibria (Fig.~\ref{fig:eqpayoff}(b)). These two cases are mutually exclusive. An analytical expression for the
boundary between the zones where they exist can also be obtained as detailed 
in Appendix~\ref{app:BoundNe}.


 Returning to the broader picture, the Nash equilibrium analysis of the double auction market choice game clearly shows that there is \emph{potential} for segregation: as illustrated in Fig.~\ref{fig:diff_ne}, heterogeneous pure strategy distributions have two peaks that indicate players within a class separating into two distinct subpopulations playing opposite pure strategies. Heterogeneous mixed strategies can similarly have two or more peaks. This emergence of segregation shows that the observations of segregation in a previous study of EWA learning~\cite{aloric2015emergence} were not based on purely dynamical effects. We also find qualitatively similar trends, e.g.\ the equilibria where both classes of players can be segregated (potentially heterogeneous) are most prevalent in Fig.~\ref{fig:phidiag} when the two markets are identical ($\theta_1=0.5$), showing that segregation is not a trivial consequence of differences between markets.

However, the Nash equilibrium conditions only identify the means of the strategy distributions $\phi^{(1)}$ and $\phi^{(2)}$. As we saw, this means for a potentially heterogeneous (or potentially partially heterogeneous) equilibrium that we cannot decide whether the underlying strategy distribution is homogeneous (mixed) or heterogeneous, nor do we know whether a heterogeneous mixed strategy distribution would actually have two distinct peaks as required for the concept of segregation to make sense.
We therefore study next under what conditions EWA learning \emph{dynamics} as defined in Sec.~\ref{sec:model} reaches as its steady state a Nash equilibrium of our system. Once this connection is established, we ask 
which particular Nash equilibria are selected as possible steady states of EWA learning. Put differently, does the learning dynamics break the indeterminacy of the Nash equilibrium conditions?

\section{EWA learning in double auction markets}
\label{sec:learn}
In this section, we study the steady states of the EWA learning
dynamics defined in Sec.~\ref{sec:model} in a game with a large number
of players. We are interested in particular when different types of steady state strategy distributions, 
as sketched in Fig.~\ref{fig:diff_ne}, can occur.

\begin{figure}[t!!]
  \centering
  \includegraphics[scale = 0.5]{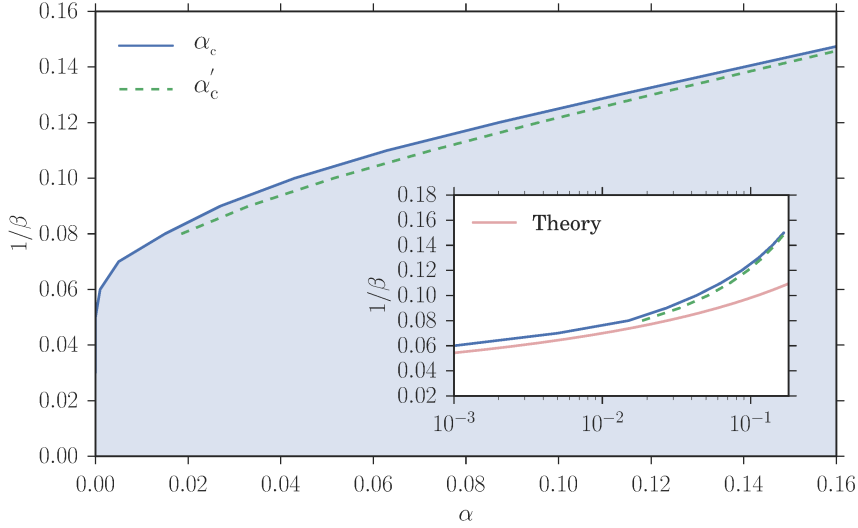}
  \caption{EWA learning phase diagram. The blue zone shows the region of the $(\alpha,1/\beta)$-plane where the steady state strategy distribution of each of the two classes of agents is heterogeneous. Elsewhere, including in particular on the line  $\alpha = 0$, the strategy distribution is homogeneous. The blue line shows the threshold $\alpha_c$
    where the distribution switches from homogeneous to heterogeneous mixed. As $\alpha$ is increased further beyond a threshold $\alpha_c'$ (dashed green line), the strategy distribution becomes heterogeneous pure. The market and trader parameters for this diagram are $\theta_1 = 1 - \theta_2 = 0.3$ and $\pb^{(1)} = 1 - \pb^{(2)} = 0.2$. Inset: Threshold curves plotted with a logarithmic $\alpha$-axis. The red line shows the  exponential  dependence of the characteristic values of $\alpha$ on $\beta$ (with an arbitrary prefactor) that is expected from the theoretical considerations in Appendix~\ref{app:scalingbeta}.
  }
  \label{fig:alphac}
\end{figure}

We argue in Sec.~\ref{sec:assum} that one expects the steady state of the EWA learning
dynamics to approach a Nash equilibrium of the model described in
Sec.~\ref{sec:mf} in the joint limit where the fictitious play
coefficient $\alpha\to 0$, the intensity of choice $\beta\to\infty$ and the inverse memory length $r\to 0$. In principle our task is thus to find the 
steady state of EWA learning and then to take this joint limit. It turns out, however, that this is far from trivial. The reason is shown by the phase diagram in
Fig.~\ref{fig:alphac}, where the limit $r\to 0$ has already been taken. What is notable is that there are different regions in the phase diagram where the steady state strategy distributions are homogeneous and heterogeneous, respectively. The Nash equilibrium limit point
$(\alpha,1/\beta) = (0,0)$ can be approached along paths within either of these regions, which means there will be several possible limiting strategy distributions of EWA learning, and it is these that we will want to identify. 
Note that we focus generally on system parameters where potentially heterogeneous Nash equilibria exist (see Fig.~\ref{fig:phidiag}), for which the EWA learning phase diagram has the generic structure of Fig~\ref{fig:alphac}.

We introduce in 
Sec.~\ref{sec:KM} the Kramers-Moyal expansion for the EWA learning dynamics on which the rest of the analysis is based. In particular, we study homogeneous and heterogeneous distributions of preferences
in Sec.~\ref{sec:homogeneous}  
and Sec.~\ref{sec:heterogeneous}, respectively, and analyse how they approach Nash equilibria in the relevant limit. The large deviation methods we deploy for the heterogeneous case are described separately in Sec.~\ref{sec:method}.

As before we choose to concentrate on settings with symmetric market biases and buy/sell preferences, and within those on steady states of the EWA learning dynamics that also have
symmetric aggregates
$\bar{p}^{(1)} = 1 - \bar{p}^{(2)}$. This captures the dominant steady states, simplifies the numerical analysis (see Sec.~\ref{sec:method}) and also makes it easier to illustrate the concepts. In the graphical representation of Fig.~\ref{fig:eqpayoff}, the steady states we are considering lie on the diagonal from top left to bottom right (dotted line).

\subsection{Nash Equilibria as limits of EWA learning}
\label{sec:assum}
In the game theoretical study of Sec.~\ref{sec:mf}, we considered  a large game ($N \to \infty$).
 The Nash equilibria we studied assume implicitly (i) that each player is able
to evaluate his expected payoff (\emph{full information assumption}),
(ii) that this evaluation averages appropriately over all stochastic effects (\emph{no fluctuation assumptions})
and (iii) that the players always choose the action with the highest payoff
(\emph{best response assumption}). One therefore expects a learning dynamics that verifies these same assumptions to converge to one of the Nash
equilibria we characterized in Sec.~\ref{sec:mf}.

We now consider when the above assumptions hold for EWA learning
dynamics. If we want the players' 
attractions to be accurate estimates of the payoffs for the corresponding
action (assumption (i)) we require
$\alpha \to 0$ to ensure that the attractions to actions that are not played do not decrease over time. To average over payoff fluctuations (assumption (ii)) we further need to work in the large memory limit $r \to 0$. To see this, note that in each training round the players' attractions are modified only by an amount of order $r$. For small $r$, attractions therefore change substantially after $\sim 1/r$ training rounds. This means the players effectively average the payoffs over many trading rounds that take place while their attractions and hence their strategies remain fixed, and in the limit obtain the correct expected payoffs \cite{sato2003coupled}. 
Finally, a large intensity of choice
($\beta \to \infty$) ensures that players best respond to their
attractions, so that EWA learning in that limit also verifies assumption (iii).

\subsection{Kramers-Moyal expansion for $r\to 0$}
\label{sec:KM}

Of the three limits identified above we take first the large memory
limit $r\to 0$. In this limit---and the large system limit
$N\to\infty$, which we always assume---the dynamics of EWA learning
can be described by a (nonlinear) Fokker-Planck
equation~\cite{aloric2015emergence}. This is derived by a
Kramers-Moyal expansion truncated at the second order; we defer the
details to Appendix~\ref{app:kmexp}. Denoting by
$\mathbb{P}(\mathbf{A}^{(c)},t)$ the distribution of attractions of
traders from class $c$, where
$\mathbf{A}^{(c)} = (A^{(c)}_1,A^{(c)}_2)$ is a vector gathering the
attractions towards market 1 and 2, the Fokker-Planck equation
describing the time evolution of this distribution
is
\begin{eqnarray} \label{eq:kmexp}
  \partial_t \mathbb{P}(\mathbf{A}^{(c)},t) &=& -\sum_{m=1}^2 \partial_{A^{(c)}_m}\left[\mu^{(c)}_m(\mathbf{A}^{(c)},\bar{p}^{(1)},\bar{p}^{(2)})\mathbb{P}(\mathbf{A}^{(c)},t)\right]\notag\\
                                            &&{}+\frac{r}{2} \sum_{m,m'=1}^{2}\partial_{A^{(c)}_m} \partial_{A^{(c)}_{m'}} \left[\Sigma^{(c)}_{m {m'}}(\mathbf{A}^{(c)},\bar{p}^{(1)},\bar{p}^{(2)})\mathbb{P}(\mathbf{A}^{(c)},t)\right]
\end{eqnarray}
Here time $t=rn$ is a rescaled version of the number of trading rounds
$n$, while $\bar{p}^{(1)}$ and $\bar{p}^{(2)}$ are the average
fractions of traders from class 1 (resp.\ class 2) choosing to go to
the first market. These fractions are obtained simply by averaging the
probability of choosing market 1 as defined in \eqref{eq:ptrade} over
the relevant distribution of attractions:
\begin{align}
  \bar{p}^{(c)} = \int {\rm d} \mathbf{A}^{(c)} \,\mathbb{P}(\mathbf{A}^{(c)},t) \sigma_\beta (A^{(c)}_1 - A^{(c)}_2)
  \label{eq:pbar_from_P}
\end{align}
Formally, $\bar{p}^{(1)}$ and $\bar{p}^{(2)}$ are therefore
functionals of the probability distributions
$\mathbb{P}(\mathbf{A}^{(c)},t)$
It is this dependence that makes the Fokker-Planck equation nonlinear,
and couples the dynamics of the attraction distributions in class 1
and 2.

At fixed values of $\bar{p}^{(1)}$ and $\bar{p}^{(2)}$, the
Fokker-Planck equation~\eqref{eq:kmexp} describes for each class the
Langevin dynamics of the attraction vector $\mathbf{A}^{(c)}$ of a
\emph{single agent}, with deterministic drift vector $\mu_m^{(c)}$ and
(multiplicative) white noise with covariance matrix
$r \Sigma_{mm'}^{(c)}$. The form of the drift follows directly from
the original EWA dynamics~\eqref{eq:EWA_dynamics} (see Appendix~\ref{app:kmexp})
\begin{equation}
  \mu_1^{(c)}(\mathbf{A}^{(c)},\bar{p}^{(1)},\bar{p}^{(2)}) = \left[\mathcal{P}^{(c)}_1(f_1(\bar{p}^{(1)},\bar{p}^{(2)})) -  A_1^{(c)}\right] \sigma_\beta(A_1^{(c)}- A_2^{(c)}) -\alpha A_1^{(c)} \left[1-\sigma_\beta(A_1^{(c)}- A_2^{(c)})\right]
  \label{eq:drift}
\end{equation}
The first term describes the change in the attraction to market 1 (in
square brackets), weighted with the probability of the agent choosing
that market. The second term corresponds to the opposite case where
the agent chooses market 2.

The Fokker-Planck equation~\eqref{eq:kmexp} is of course impossible to
solve in closed form in general. A special case is the limit $r\to 0$,
assuming the population is initially homogeneous, \ie\ a
delta-distribution.
Homogeneity is then maintained over time for $r=0$, where the dynamics
is deterministic, and Eq.~\eqref{eq:kmexp} gives for the time
evolution of the locations of the peaks of the attraction distributions
the
equations
\begin{equation}
  \partial_t A_m^{(c)} =  \mu^{(c)}_m(\mathbf{A}^{(c)},\bar{p}^{(1)},\bar{p}^{(2)})
  \label{eq:detdyn}
\end{equation}
Together with
\begin{equation}
  \bar{p}^{(c)}(t) = \sigma_\beta(A_1^{(c)}(t)- A_2^{(c)}(t))
  \label{eq:detdynaggr}
\end{equation}
one then has a system of nonlinear differential equations that is
straightforward to solve numerically.  We call this the
\emph{homogeneous populations dynamics}, where the population changes
over time but remains homogeneous.

For nonzero $r$, analysing the EWA Fokker-Planck equation becomes more
difficult because the attraction distributions broadens and can indeed
develop multiple peaks. As we are primarily interested in long-time
steady states, we focus on this somewhat simpler case. The task at
hand here is a self-consistency problem: find a set of aggregates
$\bar{p}^{(1)}$, $\bar{p}^{(2)}$ for which the steady state solution
of the Fokker-Planck equation, when inserted into
\eqref{eq:pbar_from_P}, gives back the original aggregates. If we call
$\tilde{p}^{(c)}(\bar{p}^{(1)},\bar{p}^{(2)})$ the aggregates
calculated from the steady state solution, the self-consistency
equations are simply
$\tilde{p}^{(c)}(\bar{p}^{(1)},\bar{p}^{(2)})=\bar{p}^{(c)}$. 
\begin{figure}[h! ]
  \centering
  \includegraphics[scale=0.55]{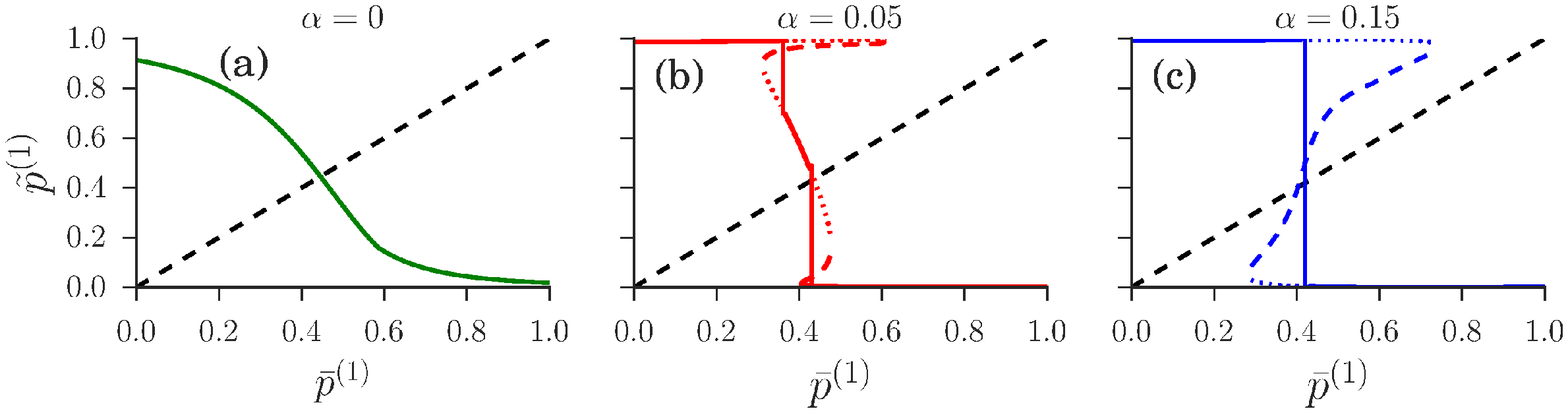}
  \caption{New aggregate $\tilde{p}^{(1)}$ calculated from steady state of single agent dynamics at  ``old'' aggregate value $\bar{p}^{(1)}$ (for $r\to 0$).
 Steady states are peaked around stable fixed points (solid/dotted), which are connected by unstable fixed points (dashed). 
In (a) only one such peak exists for any $\bar{p}^{(1)}$. The physical steady state is found from the self-consistency requirement $\tilde{p}^{(1)}=\bar{p}^{(1)}$ (dot-dashed line).
In (b,c) there are steady states with up to three peaks, but generically all but one have a weight exponentially suppressed in $1/r$ so that $\tilde{p}^{(1)}(\bar{p}^{(1)})$ (solid line) follows the curve for a single fixed point.  
At specific aggregate values the dominant peak switches and two peaks can coexist (vertical solid lines).
In (b) there are two such transitions; in (c) the middle fixed point from (b) has disappeared and there is only one transition, between branches of $\tilde{p}^{(1)}$ that are close to 0 and 1.
In (b,c) the intersection with the diagonal is \emph{at} a switch, giving a heterogeneous steady state with two peaks of comparable weight. Market and trader parameters for this figure are as in Fig.~\ref{fig:alphac}; intensity of choice $\beta= 1/0.1$.
    }
  \label{fig:selfcon}
\end{figure}
\subsection{Steady state of EWA Fokker-Planck equation}
\label{sec:steady_state}

The remaining challenge is now to determine, for small $r$, the steady state solution of the Fokker-Planck equation for given aggregates $\bar{p}^{(1)},\bar{p}^{(2)}$. As explained above, we can think of this as the steady state distribution for the dynamics of a single agent, given a fixed state of the population. In the limit $r\to 0$ this dynamics is almost deterministic so that the agent will spend almost all of her/his time near the stable fixed points of the drift $\mu^{(c)}_m$. Accordingly, $\mathbb{P}(\mathbf{A}^{(c)})$ will be peaked near these points, with the peak width being of the order of the standard deviation of the Langevin noise, \ie\ $O(\sqrt{r})$.

For aggregate values where there is only one stable single agent fixed point, $\mathbb{P}(\mathbf{A}^{(c)})$ becomes a delta-distribution centred at that point for $r\to 0$, so we have a steady state with a homogeneous distribution of attractions and hence strategies. The self-consistency condition for such a steady state is then simply 
the stationarity condition for the homogeneous population dynamics (\ref{eq:detdyn}) together with (\ref{eq:detdynaggr}). The graphical solution of this condition is illustrated in Fig.~\ref{fig:selfcon}(a).

When there are multiple stable single agent fixed points, $\mathbb{P}(\mathbf{A}^{(c)})$ for $r\to 0$ will become a sum of delta-distributions at these points. The remaining task is then to find the \emph{weight} of each of these peaks. 
We explain how to use \emph{large deviation} methods for this purpose in Sec.~\ref{sec:method}. The idea is that the peak weights are determined by the balance of fluxes of agents transitioning from one peak to another. For small $r$, the dominant $r$-dependence of these fluxes comes from exponential factors of the form $\exp(-\mathcal{S}/r)$. Fluxes can then balance for $r\to 0$ only when the ``action'' $\mathcal{S}$, which represents an effective activation barrier, is the same for the transition from one peak to the other as for the reverse transition. This condition, which is represented schematically in Fig.~\ref{fig:1}, allows one to determine the aggregate values where multiple peaks can coexist in $\mathbb{P}(\mathbf{A})$. 
At these aggregate values the steady state solution switches between two single peaked solutions. This switch happens within an aggregate value range of $O(r)$ that vanishes as $r\to 0$, giving vertical sections in the plot of $\tilde{p}^{(c)}$ versus $\bar{p}^{(c)}$ as shown in Fig.~\ref{fig:selfcon}(b).
If the intersection with the diagonal $\tilde{p}^{(c)}=\bar{p}^{(c)}$ occurs in one of these vertical sections, as in the example in Fig.~\ref{fig:selfcon}(b),
the actual peak weights can be determined indirectly from the fact that the appropriate weighted combination of the $\tilde{p}^{(c)}$ from the single peaks must give $\bar{p}^{(c)}$.
Note that one can show generally (see Appendix~\ref{app:scalingbeta}) that in each agent class there can be at most three stable fixed points, so that each $\mathbb{P}(\mathbf{A}^{(c)})$ can have at most three peaks. By choosing an appropriate aggregate value, at most two of these peaks can be made to have finite weight for $r\to 0$. Obtaining three peaks with finite weight requires one to tune $\alpha$ to $\alpha_c'$ at given $\beta$, giving the dashed green phase boundary in Fig.~\ref{fig:alphac}. Intuitively, at $\alpha_c'$ the two transitions in Fig.~\ref{fig:selfcon}(b) have moved horizontally so that they occur at the same aggregate value.

We will next study the homogeneous steady states of EWA learning dynamics.
Given the structure of the phase diagram that we anticipated in Fig.~\ref{fig:alphac}, the easiest way to ensure that steady states are homogeneous in the Nash equilibrium limit is to take $\alpha =0$.

\subsection{Homogeneous attraction distributions}
\label{sec:homogeneous}

\subsubsection*{Kramers-Moyal expansion for $\alpha = 0$}

We saw above that the dynamics of a homogeneous distributions of agents within each class is described, for $r\to 0$ by (\ref{eq:detdyn},\ref{eq:detdynaggr}). 
In steady state the right-hand side of \eqref{eq:detdyn} needs to vanish, hence using $\alpha=0$ in (\ref{eq:drift}) and its analogue for $m=2$ one has
\begin{align}
0 &=  [\mathcal{P}^{(c)}_1(f_1(\bar{p}^{(1)},\bar{p}^{(2)})) -  A_1^{(c)}] \sigma_\beta(A_1^{(c)}- A_2^{(c)})\\
0 &=  [\mathcal{P}^{(c)}_2(f_2(\bar{p}^{(1)},\bar{p}^{(2)})) -  A_2^{(c)}] \sigma_\beta(A_2^{(c)}- A_1^{(c)})
\label{eq:steady_state}
\end{align}
Here the aggregates on which $f_1$ and $f_2$ depend are given by
$\bar{p}^{(c)} = \sigma_\beta(A_1^{(c)}- A_2^{(c)})$.  In \eqref{eq:steady_state}, 
$\sigmoid(A_1^{(c)} - A_{2}^{(c)})$ cannot vanish at any finite $\beta$, so the condition for a homogeneous state is simply
\begin{align}
  \mathcal{P}^{(c)}_m(f_1(\bar{p}^{(1)},\bar{p}^{(2)}))  - A_m^{(c)} = 0 \label{eq:cond2}
\end{align}
which needs to be verified for each market $m$ and each class $c$.
This means that for each player, in the steady state of the EWA
learning dynamics, the respective attraction to each market equals the expected payoff there. The aggregates calculated from the steady state are therefore
\begin{equation}
\tilde{p}^{(c)}(\bar{p}^{(1)},\bar{p}^{(2)}) = \sigma_\beta\left(
  \mathcal{P}^{(c)}_1(f_1(\bar{p}^{(1)},\bar{p}^{(2)}))  - 
  \mathcal{P}^{(c)}_2(f_2(\bar{p}^{(1)},\bar{p}^{(2)}))\right)
\end{equation}

We now need to solve the self-consistency condition $\tilde{p}^{(c)}=\bar{p}^{(c)}$as explained in Sec.~\ref{sec:steady_state}. This can be visualized most easily if we focus on symmetric situations where $\bar{p}^{(1)}=1-\bar{p}^{(2)}$: one just has to plot the curve $\sigma_\beta(\mathcal{P}^{(1)}_1-\mathcal{P}^{(1)}_2)$ vs $\bar{p}^{(1)}$ and intersect it with the diagonal, as shown in Fig.~\ref{fig:selfcon}(a).

To retrieve EWA steady states corresponding to Nash equilibria, we need to consider the limit $\beta\to\infty$ of high intensity of choice. Then $\sigma_\beta(\mathcal{P}^{(1)}_1-\mathcal{P}^{(1)}_2)$ approaches one if the payoff at the first market $\mathcal{P}^{(1)}_1$ is larger than at the second, otherwise zero. Where the payoffs are equal, a step in the curve results, which will always produce an intersection and hence a self-consistent solution. Because of the payoff equality, such solutions correspond exactly to potentially heterogeneous Nash equilibria (see Eq.~\eqref{eq:eqpayoffcond} in Sec.~\ref{sec:mf}). Here this type of Nash equilibrium is realized in a \emph{homogeneous mixed} form: all players from class 1 play the same strategy, choosing market 1 with probability $\bar{p}^{(1)}$.

If the payoffs $\mathcal{P}^{(1)}_1$ and $\mathcal{P}^{(1)}_2)$ are different across the entire range of $\bar{p}^{(1)}$, we have a different scenario: assuming 
$\mathcal{P}^{(1)}_1 > \mathcal{P}^{(1)}_2$ for definiteness, $\sigma_\beta(\mathcal{P}^{(1)}_1-\mathcal{P}^{(1)}_2)$ tends to one for $\beta\to\infty$, hence the only self-consistent solution is $\bar{p}^{(1)}=1$. 

This corresponds to a \emph{homogeneous pure} Nash equilibrium, with---because of the assumed symmetry---the two classes of players trading at different markets.

\begin{figure}[t!!]
  \centering
  \includegraphics[scale = 0.6]{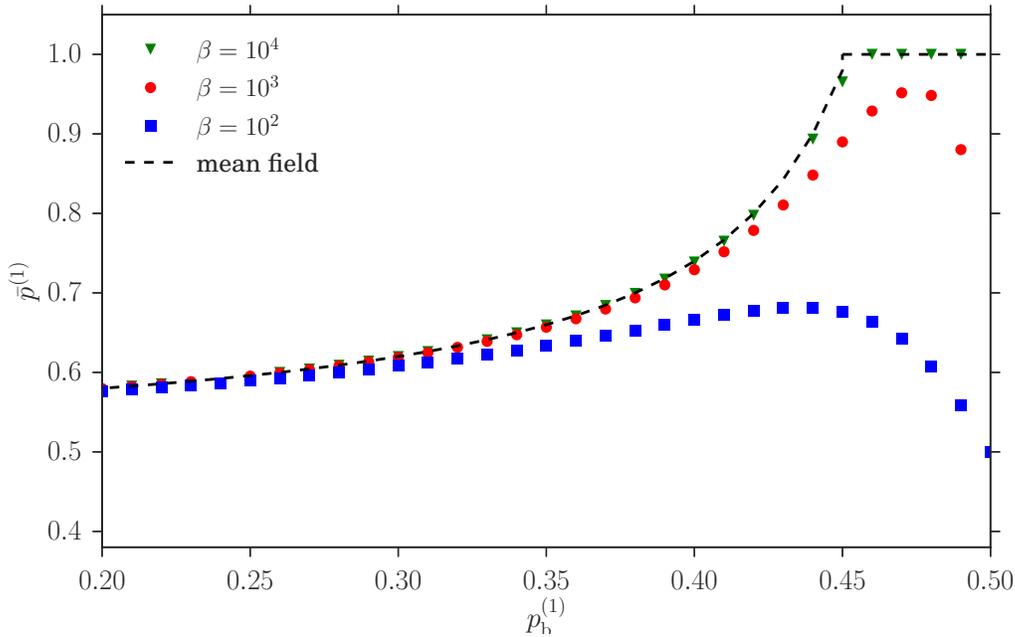}
  \caption{Comparison between mean field Nash equilibria (continuous
    lines) and homogeneous steady states of EWA learning (symbols) for three different values of the intensity of choice $\beta$. The market biases are
    $\theta_1 = 1- \theta_2 = 0.3$ and the buying probabilities
    $\pb^{(1)} = 1 - \pb^{(2)} = \pb$. Shown is $\bar{p}^{(1)}$, the fraction of traders from the first class going to the first market, versus $\pb^{(1)}$. 
  }
  \label{fig:mf_vs_dyn}
\end{figure}

To show the approach to the large $\beta$-limit, we show in Fig.~\ref{fig:mf_vs_dyn} numerically determined values of $\bar{p}^{(1)}$, the fraction of traders from the first class
going to the first market in the steady state of EWA learning. The results for three different $\beta$ are compared to the values of $\bar{p}^{(1)}$ determined from the mean field Nash equilibrium condition, which as we saw leads to the
two payoff
equalities~\eqref{eq:eqpayoffcond}. As expected, as $\beta$ gets larger, the
aggregate $\bar{p}^{(1)}$ gets closer to its Nash equilibrium value, confirming our reasoning above. Note around $\pb^{(1)}=0.45$
we transition from the situation in Fig.~\ref{fig:eqpayoff}(a), where the Nash equilibrium and the corresponding steady state are of homogeneous mixed type (green triangle in the figure), to the homogeneous pure state (orange square) in Fig.~\ref{fig:eqpayoff}(b).

So far our main conclusion is that steady states of EWA learning can give \emph{homogeneous mixed} realizations of the potentially heterogeneous Nash equilibria we had identified in Sec.~\ref{sec:mf}: even though the equilibrium could be heterogeneous, the dynamics generates a homogeneous steady state with the same aggregates where all players use the same mixed strategy. This happens if we consider the limit of the dynamics for $\beta\to\infty$ at $\alpha=0$. One would expect from the phase diagram in Fig.~\ref{fig:alphac}
that the same steady state is obtained if we move the path of approach towards $(\alpha,1/\beta)=(0,0)$ slightly away from the vertical axis, \ie\ if $\alpha$ is nonzero but goes to zero sufficiently fast as $\beta$ grows.
We show in Appendix~\ref{app:scalingbeta} that this is true if the decay of $\alpha$ is exponential, 
$\alpha_c \sim \exp(-\mbox{const}\cdot\beta)$: if the constant in the exponent is large enough, the attraction distributions remain homogeneous and attractions again become equal to payoffs for $\beta\to\infty$.

\subsection{Heterogeneous attraction distributions}
\label{sec:heterogeneous}

We investigate in this section steady states of EWA learning where the attraction distributions of traders are multimodal (heterogeneous) rather than unimodal. As explained in Sec.~\ref{sec:steady_state}, for $r\to0$ the modes become sharp peaks so that unimodal distributions become homogeneous. We have investigated the latter case so far, but heterogeneous steady states should also exist. Indeed, it was shown in
\cite{aloric2015emergence} using multi-agent simulations as well as
theoretical studies of the Kramers-Moyal expansion detailed in
Sec.~\ref{sec:KM} that for high enough intensity of choice $\beta$
the distribution of attractions undergoes a transition from
homogeneous to heterogeneous. We therefore expect to find heterogeneous steady states of EWA learning more generally for large $\beta$ and $\alpha$ not too small. We confirm this expectation in this section, where we also find surprising transitions between different types of heterogeneous steady states.

\subsubsection*{Difference between the case of homogeneous and heterogeneous attraction distributions}
In~\cite{aloric2015emergence}, Alori\'c \etal{} describe a method to
obtain the critical $\alpha$ at which the attraction distributions
of the traders in the two classes become heterogeneous.
One assumes initially that the distributions are
homogeneous and determines a self-consistent assignment of the aggregates $\bar{p}^{(1)}$, $\bar{p}^{(2)}$ on this basis. One then checks whether the single agent dynamics for these aggregate values
has one fixed point, producing a homogeneous distribution of attractions, or two or more (stable) fixed points, giving a heterogeneous distribution with peaks at these locations in attraction space.
What this method leaves open, however, is what the weights of these peaks are and in particular whether they remain nonzero in the large memory limit $r\to 0$.
This is the task we tackle using large deviation methods, as summarized in Sec.~\ref{sec:steady_state} above and described in more detail in~\ref{sec:method}.

\begin{figure}[t!!]
  \centering
  \includegraphics[scale = 0.4]{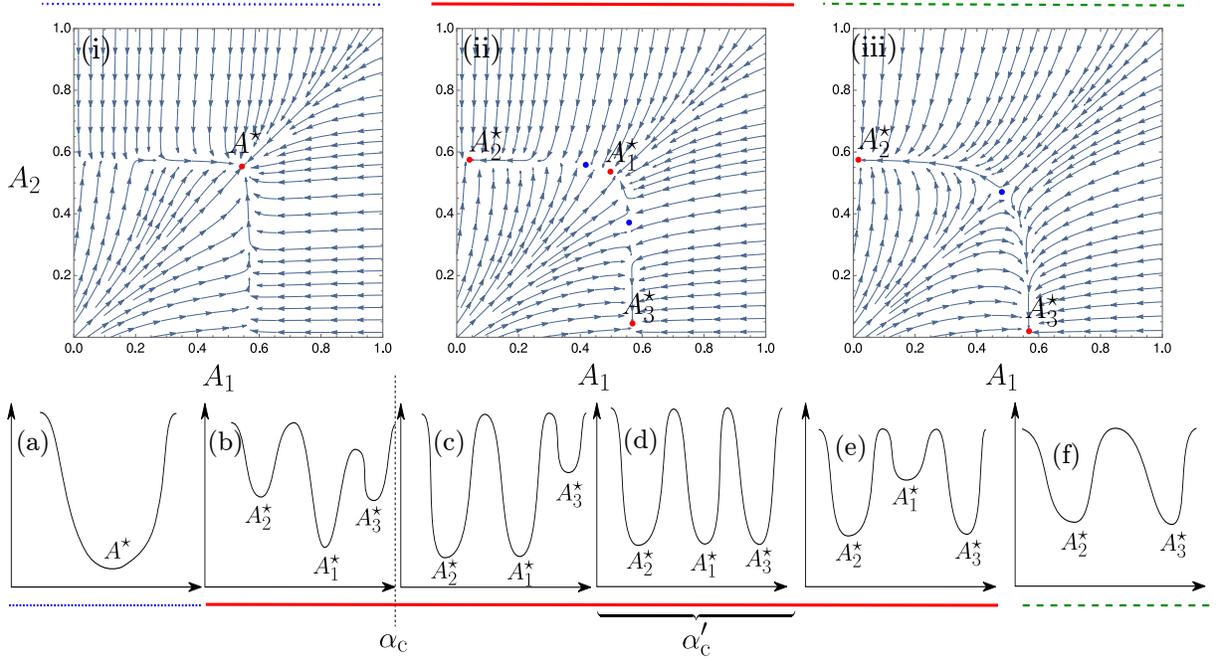}
  \caption{(i-iii) Flow diagrams of the single agent dynamics for increasing
    $\alpha$. The points represent the stable
    (red) and unstable 
(blue) fixed points of the dynamics. The
    potentials in the bottom row represent schematically in 1-D
    the arrangement of fixed points (stable = potential minimum, unstable = potential maximum).
    Attraction distributions are peaked around stable fixed points; in the 1-D representation, the lowest minima indicate peaks with weights of order unity as $r\to 0$, while higher-lying (metastable) minima correspond to peaks that become exponentially suppressed.
    For $\alpha < \alpha_c$
    , the aggregates of
    the single agent dynamics are deduced by self consistency from the
    only stable fixed point of the dynamics (panels (a) and (b)), while for larger $\alpha$ the aggregates are
    chosen such that the transition rates between the stable fixed
    points ($\boldsymbol{A}^\star_1$ and $\boldsymbol{A}^\star_2$ for 
    $\alpha_c<\alpha < \alpha_{\rm c}'$, panels (c) and (d);
$\boldsymbol{A}^\star_1$ and $\boldsymbol{A}^\star_3$ for 
$\alpha > \alpha_{\rm c}'$, panels (e) and (f)) are of the same order. Plots were produced with symmetric market biases $\theta_1 = 1- \theta_2 = 0.3$ and probability of buying $\pb^{(1)} = 1 - \pb^{(2)}= 0.2$ and intensity of choice $\beta = 1/0.11$.
}
  \label{fig:5fp}
\end{figure}

\subsubsection*{Transition from one to two to three stable fixed points}

We next explore the different fixed point
structures of the single agent dynamics as a function of the
fictitious play parameter $\alpha$, for fixed large intensity of choice
$\beta$. In principle at each $\alpha$ the aggregates $\bar{p}^{(c)}_1$ , $\bar{p}^{(c)}_2$ need to be determined from self-consistency but from the experience with the homogeneous solutions we expect that as long as $\alpha$ is small enough and $\beta$ large enough, the self-consistent aggregate values will be close to their Nash equilibrium values. To leading order one can therefore think of varying $\alpha$ at fixed aggregates. As before we also rely on the assumption
that the memory of the traders is large ($r \to 0$); the finite memory
case will be investigated below using numerical simulations.

When the fictitious play coefficient $\alpha$ is small enough, the
single agent dynamics has a single stable fixed point
$\boldsymbol{A}^\star_1$ (see Appendix~\ref{app:scalingbeta} and Fig.~\ref{fig:5fp}(i)) and so for $r\to 0$ the distribution of attractions is a
$\delta$-peak
at this point as shown in
Fig.~\ref{fig:diff_ne}(a). As $\alpha$ increases then
as shown in Fig.~\ref{fig:5fp}(b) two new stable fixed points
$\boldsymbol{A}^\star_2$ and $\boldsymbol{A}^\star_3$ appear, first one and then the other. But the 
distribution of attractions is still delta peaked around the original fixed point because in the limit $r\to 0$ the other fixed points are exponentially suppressed in $1/r$: they are in this sense metastable.

The first phase transition arises
at a critical value of $\alpha$, $\alpha_c$, where one of the
metastable point becomes stable; in
Fig.~\ref{fig:5fp} this is $\boldsymbol{A}^\star_2$. In this case, the attraction distribution is
composed of two $\delta$-peaks located at these two stable fixed points
of the single agent dynamics (see Fig.~\ref{fig:diff_ne}(a,b) for an
example projected onto one direction in attraction space). The transition occurs because the actions (see Sec.~\ref{sec:method}) for single agents to move  from one stable fixed point to the other and for the reverse move become equal.

This ensures that the fluxes of agents between the two stable fixed points are of the same order of magnitude in both directions, and hence that the two peaks in the attraction distribution can have comparable rather than exponentially different weights.

As $\alpha$ increases further, small changes to the aggregates maintain the condition of comparable flux between the two existing stable peaks. Eventually, at some $\alpha_{\rm c}'$ 
higher than $\alpha_c$,
the third fixed point also becomes stable so that the attraction distribution acquires three peaks.

Note that the weights of the three peaks cannot be fully determined at $\alpha=\alpha_{\rm c}'$: the self-consistency for $\bar{p}^{(1)}$ only gives one condition for three nonnegative peak weights that need to sum to one, so that the problem is underconstrained. This indicates that for nonzero $r$ these weights would vary continuously across a small range of $\alpha$ of order $r$.

For $\alpha>\alpha_{\rm c}'$, it is the turn of the central fixed point
$\boldsymbol{A}^\star_1$ to become metastable;  aggregate values are
determined by the equal action condition between the two outer stable fixed
points and the attraction distribution goes back to having only two $\delta$-peaks. 
Finally at even larger $\alpha$
the central metastable fixed point disappears altogether in a saddle-node bifurcation.

\subsubsection*{Game theoretical interpretation of the steady states}

We now investigate the characteristics of all the
steady states described above and compare each of them to the Nash
equilibria enumerated in Sec.~\ref{sec:mf}. When $\alpha$ is below the
critical value $\alpha_c$, all the traders within one class
randomize between the two markets, going to the first market
with the same probability. This probability is
$\sigma_\beta(A_1^{(c)}- A_2^{(c)})$ evaluated at the stable fixed
points of the single agent deterministic dynamics, which also equals $\bar{p}^{(c)}$ (see Sec.~\ref{sec:assum}).
 This
\emph{homogeneous mixed} strategy profile is plotted as the single-peaked preference distribution in
Fig.~\ref{fig:1}. 

For the opposite case of large $\alpha$, 
$\alpha > \alpha_{\rm c}'$, there are  within each class two sub-populations of traders, each of which 
corresponds to a peak of the attraction distribution as shown schematically in Fig.~\ref{fig:diff_ne}(b). Looking at Fig.~\ref{fig:5fp}(iii) and (f), one sees that at both of these peaks, the attractions to the two markets remain distinct for large $\beta$ -- the relevant fixed points are far from the 45$^\circ$ diagonal. In the limit both sub-populations will therefore play a pure strategy as $\sigma_\beta(A_1^{(c)}- A_2^{(c)})$ tends to one or zero, respectively.
This situation is shown as the preference distribution in Fig.~\ref{fig:diff_ne}(b) with two peaks around preference one and zero, representing two sub-populations of traders all choosing market 1 and 2 respectively. This steady state of EWA learning is therefore a \emph{heterogeneous pure} realization of a Nash equilibrium, as the
preferences of traders are heterogeneous, with two sub-population
playing different pure strategies.

While the two cases of homogeneous mixed and heterogeneous pure Nash equilibria are well studied in the literature \cite{Schmeidler1973,cabral1998}, we find a novel state
for $\alpha_c <\alpha <\alpha_{\rm c}{'}$. Again there are within each class two sub-populations of traders.
But now
one sub-population has attractions that become equal for large $\beta$: the corresponding fixed point lies close to the diagonal in Fig.~\ref{fig:5fp}(ii). These traders therefore play a mixed strategy and randomize between the two
markets. Overall we have a \emph{heterogeneous mixed} steady state because not all traders play pure strategies.
This is illustrated in the right panel of Fig.~\ref{fig:1}. Such
heterogeneous mixed strategy distributions have, to our
knowledge, never been reported in any study of aggregative games so it is fascinating that they are accessible by EWA learning dynamics.

Overall, we have found that potentially heterogeneous Nash equilibria can be realized as steady states of EWA learning in three different ways by appropriately taking the limits of perfect fictitious play $\alpha\to 0$ and best response $\beta\to\infty$. For small enough $\alpha<\alpha_{\rm c}(\beta)$ one obtains a homogeneous mixed equilibrium, while keeping larger $\alpha>\alpha_{\rm c}'(\beta)$ gives a heterogeneous pure equilibrium. Most interesting is the case where $\alpha$ is taken to zero in the ``corridor'' $\alpha_{\rm c}<\alpha<\alpha_{\rm c}'$, which results in a heterogeneous mixed equilibrium.

Note that the partially heterogeneous
Nash equilibria (where one class of traders splits into sub-populations while the other stays homogeneous) do not appear in the analysis above because we
restricted ourselves to studying Nash equilibria for which the aggregates are
symmetric ($\bar{p}^{(1)} = 1- \bar{p}^{(2)}$), thus ruling out
partially heterogeneous Nash equilibria.

\begin{figure}[t!!]
  \centering
  \includegraphics[scale = 0.7]{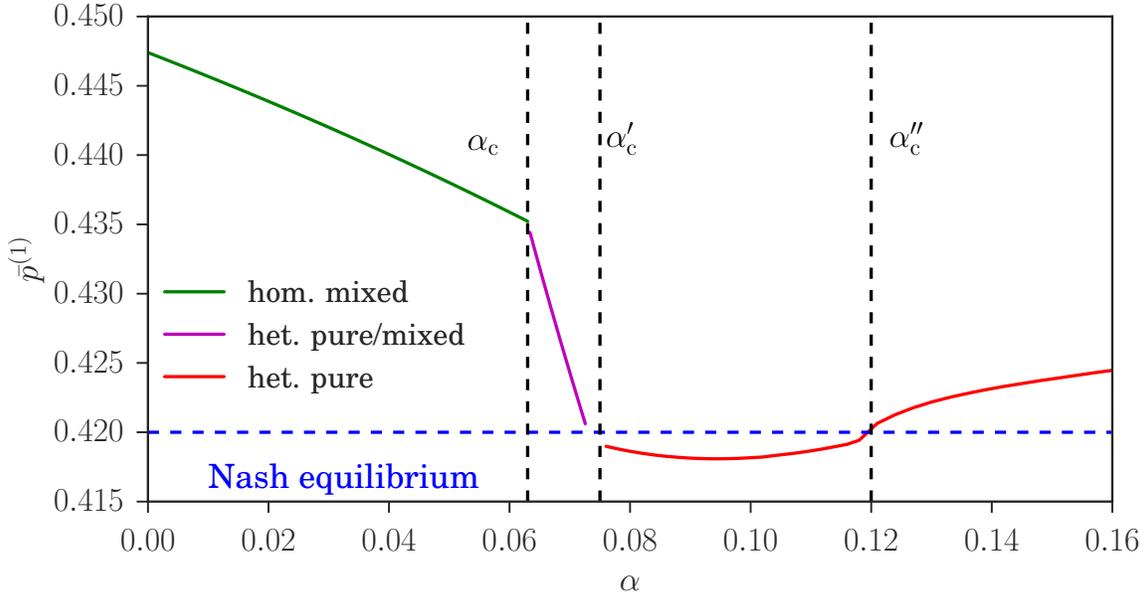}
  \caption{Fraction of traders from the first class in the first
    market, $\bar{p}_1^{(1)}$, for intensity of choice $\beta = 1/0.11$, compared with the value of $\bar{p}^{(1)}$ calculated for the corresponding potentially
heterogeneous Nash equilibrium (see Sec.~\ref{sec:mf}). Note that the deviation between the two values is small throughout. Critical values of $\alpha$ separating the different types of steady states are indicated; $\alpha_{\rm c}''$ is the value of $\alpha$ where the ``central'' fixed point representing traders playing mixed strategies disappears. Same system parameters as in Fig.~\ref{fig:5fp}.
}
  \label{fig:T11}
\end{figure}

We close this section by showing in Fig.~\ref{fig:T11} some numerical results for the aggregate $\bar{p}^{(1)}$ as a function of $\alpha$, for a fixed intensity of choice $\beta$. The values of $\alpha_{\rm c}$ and $\alpha_{\rm c}'$ are shown to indicate the transitions between the homogeneous mixed, heterogeneous mixed and heterogeneous mixed states as $\alpha$ grows. Also shown is the even larger critical value $\alpha_{\rm c}''$ at which the ``central'' fixed point (see Fig.~\ref{fig:5fp}) disappears. Note the vertical scale of the plot, which demonstrates a key point: even though $\beta=1/0.11$ is not yet very large, $\bar{p}^{(1)}$ is already quite close to the value $\bar{p}^{(1)}\approx 0.42 $ for the potentially heterogeneous Nash equilibrium as calculated using the equal payoff criterion \eqref{eq:eqpayoffcond} in Sec.~\ref{sec:mf}. 

\begin{figure}[t!!]
  \centering
  \includegraphics[scale = 0.6]{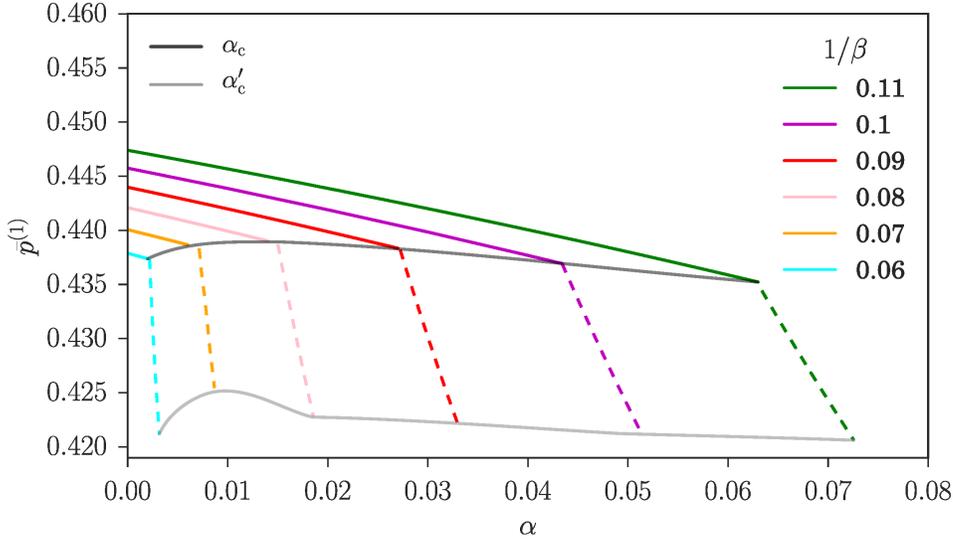}
  \caption{Fraction of players from class $1$ in the first market, $\bar{p}^{(1)}$, for different values of $\beta$. The grey lines connect the values at the two critical $\alpha$ (see Fig.~\ref{fig:5fp}) as a guide to the eye.
System parameters  as in Fig.~\ref{fig:5fp}. Note that $\bar{p}^{(1)}$ gets progressively closer to the Nash equilibrium value $\approx 0.42$ as the intensity of choice $\beta$ grows.
}
  \label{fig:algores}
\end{figure}

As we have argued this agreement should get even better as $\beta$ grows. Numerical data supporting this are shown in Fig.~\ref{fig:algores}: $\bar{p}^{(1)}$ decreases towards the Nash equilibrium value with increasing $\beta$. 
Also displayed are the critical values $\alpha_{\rm c}$ and $\alpha_{\rm c}'$, which as expected tend to zero as $\beta$ grows. It is these values that were used to produce the phase diagram in Fig.~\ref{fig:alphac}.

We note as an aside that in Fig.~\ref{fig:algores} the variation of $\bar{p}^{(1)}$ with $\alpha$ 
is rather steeper in the heterogeneous mixed
phase (between $\alpha_{\rm c}$ and $\alpha_{\rm c}'$) than in the homogeneous mixed regime. This probably reflects the change in the way the aggregates are 
determined in the two regimes: in the homogeneous-mixed phase the aggregates are obtained only by the self-consistency condition for the fixed point location, while they are fixed by the equal
flux condition in the heterogeneous mixed phase.

\subsubsection*{Test against simulations}

In this section we test the theoretical predictions obtained above in the $r \to 0$ and for infinite population size $N$ against 
agent based simulations with a finite memory ($r>0$) and finite $N$.
We are primarily interested in the steady state of the attraction distribution of the agents, but also consider its time evolution to this steady state. We continue to consider symmetric scenarios so focus on the properties of agents of class 1 throughout.
Depending on where the key parameters $\alpha$ and $\beta$ are  in the phase diagram of Fig.~\ref{fig:alphac}, one expects qualitatively different shapes for the attraction distribution resulting from the learning dynamics.
We present simulation results in each of the distinct regions of the phase diagram in Fig.~\ref{fig:alphac}.

\begin{figure}[h! ]
  \centering
  \includegraphics[scale=0.5]{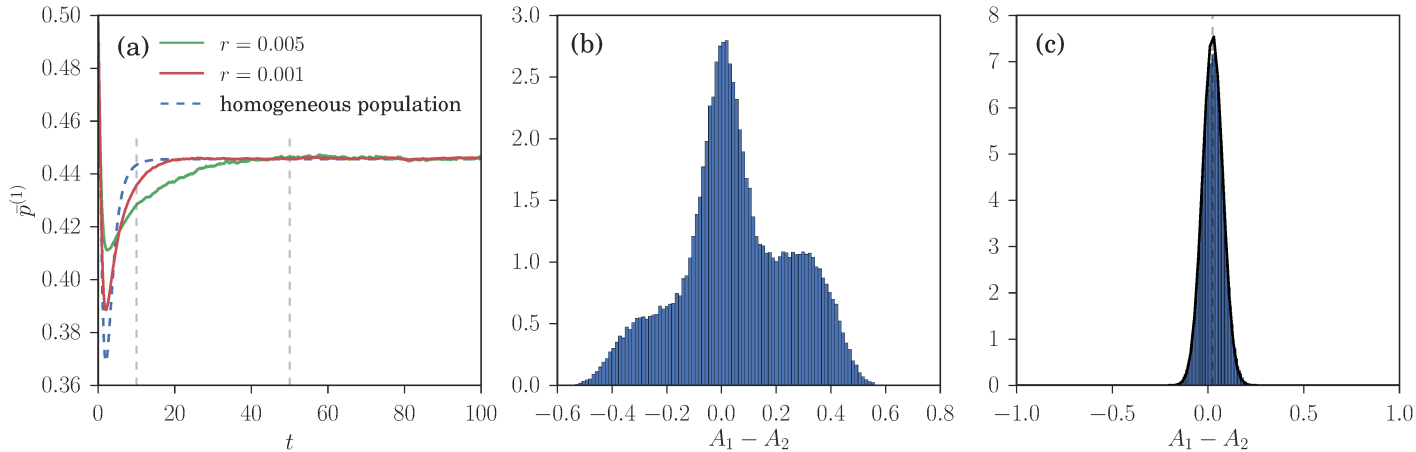}
  \caption{EWA learning dynamics at small $\alpha=0.01$. (a) Time evolution of 
$\bar{p}^{(1)}$ for 
$r=0.005$ and $r=0.001$ compared to the homogeneous population dynamics predicted for ($r \to 0$). (b,c) Distribution of attraction differences
across traders of class 1 at two times, for $r=0.005$. Black lines are theoretical predictions based on the homogeneous population dynamics and agree well at small $r$ and late times $t$ as expected (see text).
Note that for the larger $r$, the dynamics (a) and the attraction distributions   (b) deviate from the small-$r$ theory, showing a transient segregation behaviour that is the precursor of steady state segregation (see Fig.~\ref{fig:SimuMediumAlpha}~(d)) at larger $\alpha$.
 The parameters used for those simulation are $\beta = 1/0.11$, $\theta_1 = 1 - \theta_2 = 0.3$, $\pb^{(1)} = 1 - \pb^{(2)} = 0.2$, the system is composed of $20000$ traders.}
  \label{fig:SimuLowAlpha}
\end{figure}

The first zone of interest is on the far left of the phase diagram, where $\alpha$ is below the first
segregation threshold $\alpha_c$.
Here, in
the steady state of the learning dynamics, we observe in Fig.~\ref{fig:SimuLowAlpha}(c) the homogeneous
distribution of preference predicted by the theory. Looking beyond this agreement for the steady state at the time evolution,
panel~\ref{fig:SimuLowAlpha}(a) shows that for $r = 0.005$ the transient dynamics of the
aggregates is  nonetheless different from the homogeneous population
deterministic dynamics. This appears to be related to a transient
segregation effect observed in a small time window around $t = 10$
(Fig.~\ref{fig:SimuLowAlpha}(b)).
This transient segregation does not occur for lower
values of $r$ (\eg{} $r = 0.001$), where the dynamics of the aggregates is
closer to the homogeneous population dynamics (see
Fig.~\ref{fig:SimuLowAlpha}(a)).

\begin{figure}[h! ]
  \centering
  \includegraphics[scale=0.4]{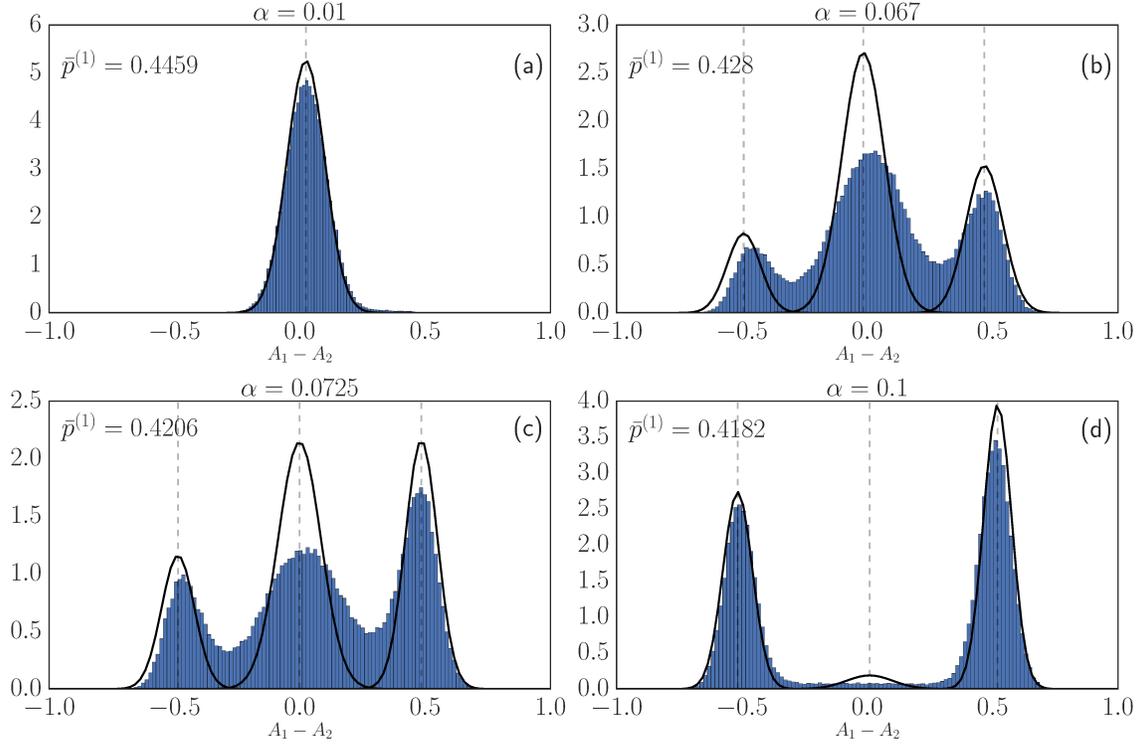}
  \caption{Steady state distribution of the attraction differences for $ r = 0.01$, and increasing values of $\alpha$; the remaining parameters are as in Fig.~\ref{fig:5fp}.
When $\alpha = 0.067$ (panel (b)), the theory predicts one outer peak on the right and one inner peak corresponding to a fraction of the population playing a mixed strategy. The simulations additionally show an outer peak on the left, which arises from the fact that the finite-$N$ system is not in a true steady state.
Panel (c) shows the situation for $\alpha = 0.0725$,  which is the critical value $\alpha_c'$ at which
we expect to see from theory three different peaks in the distribution of
attraction differences. The theoretical predictions (black curves) is a
    Gaussian mixture composed of three peaks whose mean and variance are obtained from the Kramers-Moyal expansion while their weights, which the theory cannot predict, are fitted to the data. The peak positions are in good agreement with theory while the simulations
 overestimate the variance of the peaks, again because of transient effects.
 In panel (d), for $\alpha>\alpha_c'$, there is very good agreement with theory except for a small central peak 
that for $r\to0$ is predicted to have weight zero.
This is likely to be an effect of the nonzero $r=0.01$ used in the simulations.
}
  \label{fig:SimuMediumAlpha}
  
\end{figure}

\begin{figure}[h!]
  \centering
  \includegraphics[scale=0.7]{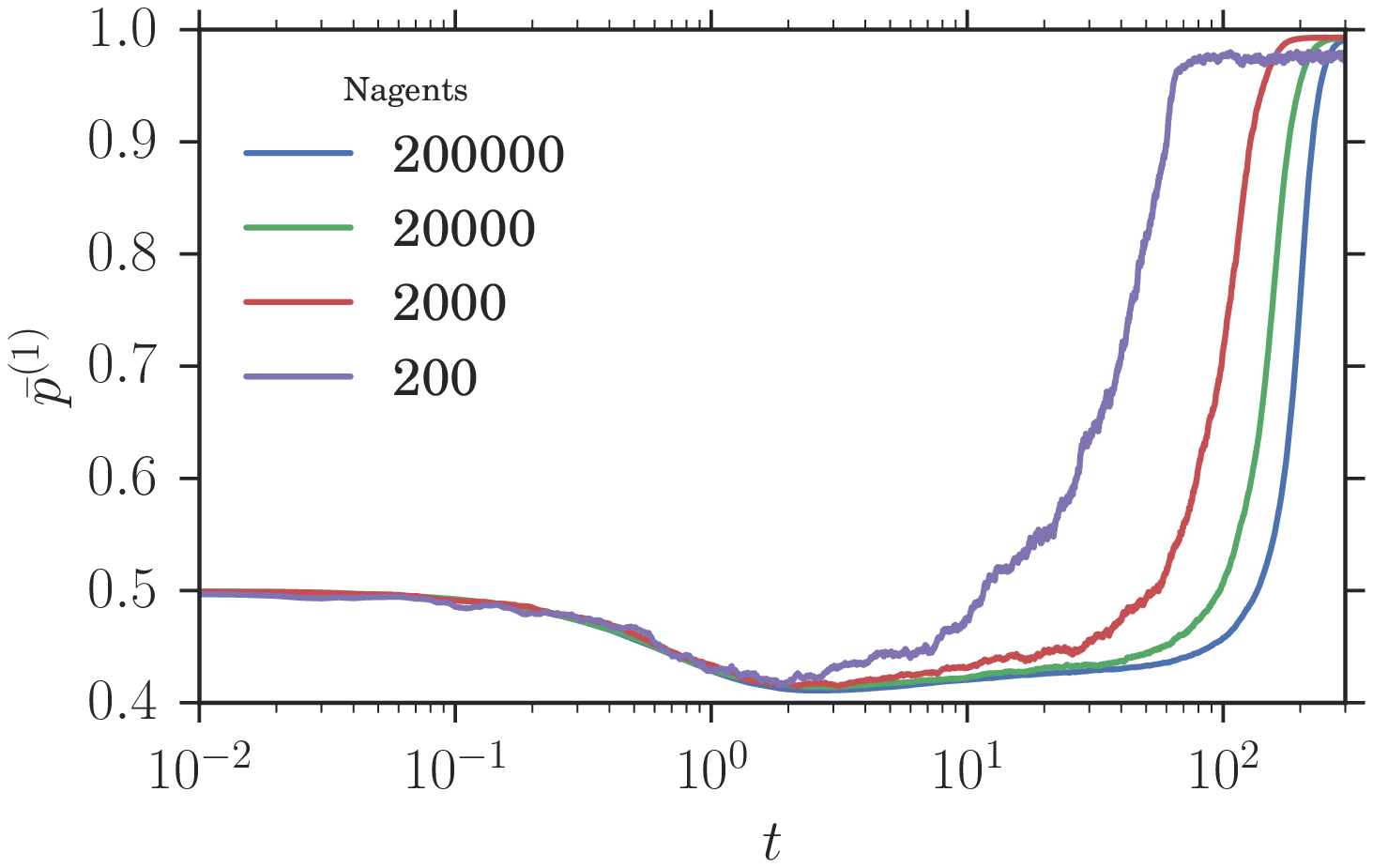}
  \caption{Time evolution of $\bar{p}^{(1)}$ for $\alpha = 0.068$,
    $r = 0.005$ and different numbers of agents $N$. Other parameters are the same as in Fig.~\ref{fig:5fp}}.
  \label{fig:numpl}
\end{figure}

When $\alpha \in [\alpha_c,\alpha_{\rm c}']$, the aggregates
relax close to their value in a Nash equilibrium around which they
fluctuate. Then, they escape from this state to reach an heterogeneous
pure Nash equilibrium. The time they remain close to the Nash
equilibria depends on the number of agents in the simulation as shown
in Fig.~\ref{fig:numpl}. The theory predicts a distribution composed
of two peaks, one peak corresponding to a sub-population playing mixed
strategies and the second one to a sub-population playing pure
strategies. The results of our simulation presented in
Fig.~\ref{fig:SimuMediumAlpha}(a) show a preference distributions
composed of three peaks, not two as the theory predicts.
On also notices that while the theoretical predictions for
the location of the peaks are consistent with the simulation results,
the width of the peaks in the simulations is larger than predicted. We believe this is because the theoretical predictions for the width of the peaks make the assumption that the
system is in its steady state. This is not strictly verified here as
the finite-$N$ system is in a transient state before relaxing to a
heterogeneous pure distribution of strategies. As $\alpha$ goes above
the second segregation threshold, $\alpha_{\rm c}'$, the dynamics initially continues to show three peaks, but in qualitative agreement with the theory the
size of the central peak diminishes rapidly, becoming negligible for large enough $\alpha$. 
The preference distributions obtained from
simulations are then consistent with the theoretical predictions as shown in
Fig~\ref{fig:SimuMediumAlpha}(d). Moreover, the aggregates stay close to
$f_1 =0.42$ and never diverge to $f_1 = 1$ or $f_1 = 0 $ (as happens for lower values of $\alpha$).

In summary, the simulations are in good qualitative accord with the predicted sequence of steady states for increasing $\alpha$: homogeneous mixed, heterogeneous mixed (outer and central peak), heterogeneous mixed (three-peaked) and finally heterogeneous pure (two outer peaks). Corrections to the theoretical predictions arise from the fact that some steady states have a lifetime that only becomes infinite for $N\to\infty$, and from the use of nonzero $r$ in the simulations.


\section{Large deviation methods}
\label{sec:method}

We describe in this section the large deviation methods we use to study heterogeneous attraction distributions in the steady state of EWA learning. As explained in Sec.~\ref{sec:learn}, steady state attraction distributions for small $r$ will be peaked around the stable fixed points of the single agent dynamics. The shape of these peaks becomes Gaussian for $r\to 0$, with a covariance matrix proportional to $r$ that is straightforward to determine. Much more difficult to find are the \emph{weights} of the peaks as these involve rare fluctuations of an agent making the transition from one peak to another. In one dimension the problem is tractable as an explicit formula for the steady state distribution of attractions can be given~\cite{aloric2015emergence}. In higher dimensions detailed balance~\cite{risken1984fokker} has a similar simplifying effect, but our single agent dynamics in the two-dimensional attraction space (for each class of agents) does not have this property.

In our approach we consider the peak weights in an attraction distribution as a result of the balance between transitions between the various peaks. We therefore need to find the rates for these transitions. To do this, note from the Kramers-Moyal expansion sketched in Sec.~\ref{sec:KM} that the single agent EWA learning is described by a Langevin equation with noise variance $O(r)$. For $r\to 0$ we are therefore looking for transition rates in a low noise limit. This allows us to use Freidlin-Wentzell theory, which deals the with large deviations of Langevin dynamics in exactly this limit~\cite{fwtheory}.

\subsubsection*{Freidlin-Wentzell theory}
We use Freidlin-Wentzell theory in the form developed in~\cite{bouchetetal,bradde2012generalized}, which generalizes, the Eyring-Kramers~\cite{kramers1940brownian} formula for the rates of noise-activated transitions to non-conservative dynamics such as our EWA learning. 
We give a brief summary of those aspects of Freidlin-Wentzell theory that we use in our numerical
application and refer to~\cite{fwtheory} for a mathematically rigorous
description
and to~\cite{bouchetetal}
for a more statistical physics-oriented summary. 

Freidlin-Wentzell theory is concerned with the transition rates between two
stable states (here $\boldsymbol{A}^\star_1$ and
$\boldsymbol{A}^\star_2$) of a non-conservative stochastic dynamics in
the low noise limit.
A general Langevin equation can be written in the form
\begin{equation}
  \label{eq:KM}
  \dot{\boldsymbol{A}}^{(c)}(t) = \boldsymbol{\mu}^{(c)}(\mathbf{A}^{(c)}(t),\bar{p}^{(1)},\bar{p}^{(2)}))+
  \sqrt{r} {\boldsymbol{\Sigma}^{(c)}}^{1/2}(\mathbf{A}^{(c)}(t),\bar{p}^{(1)},\bar{p}^{(2)})) \pmb{\xi}(t)
\end{equation}
where $\pmb{\xi}(t)$ is white noise with unit covariance matrix. 
The drift $\boldsymbol{\mu}$ and the covariance matrix
$\boldsymbol{\Sigma}$ of the noise in the Langevin equation are given in Appendix~\ref{app:kmexp} for our specific case of EWA learning, where the Langevin description results from a second order Kramers-Moyal expansion \eqref{eq:app:km}. In the generic version above we have omitted the superscript $(c)$ indicating the class of agents we are considering, as well as the dependence of drift and noise covariance on the aggregates $\bar{p}^{(1)}$ and $\bar{p}^{(2)}$.

Associated with the Langevin dynamics is an Onsager-Machlup action
$\mathcal{S}[\boldsymbol{A}]$ for any path
$\boldsymbol{A}(t)$:

\begin{equation}
  \mathcal{S}[\boldsymbol{A}] = \int_{t_1}^{t_2} \frac{1}{2} \left( \dot{\mathbf{A}}(t) -
    \boldsymbol{\mu}(\mathbf{A}(t))\right)^{T} {\boldsymbol{\Sigma}}^{-1}(\mathbf{A}(t)) \left( \dot{\mathbf{A}}(t) -
    \boldsymbol{\mu}(\mathbf{A}(t))\right) {\rm d}t
  \label{eq:action}
\end{equation}

The action determines the probability
of observing any path $[\boldsymbol{A}(t)]$ according to
\begin{equation}
 \Gamma_{1 \to 2} \sim \exp(-\mathcal{S}[\mathbf{A}]/r)
\end{equation}
where $\sim$ means that the equality is true up to a pre-factor (which depends on the time discretization used).

The main Freidlin-Wentzell result we need is that the rate $\Gamma_{1\to 2}$ for a transition from $\boldsymbol{A}^\star_1$ to
$\boldsymbol{A}^\star_2$ (\emph{forward path})
is~\cite{fwtheory,bunin2012large}
\begin{equation}
  \label{eq:fwtrate}
  \Gamma_{1 \to 2} \sim \exp(-\mathcal{S}^\star_{1\to 2}/r)
\end{equation}
where $\mathcal{S}^\star_{1\to 2}$ is the minimal action achievable by any 
paths from $\boldsymbol{A}^\star_1$ to $\boldsymbol{A}^\star_2$ in the infinite time interval $(t_1,t_2)=(-\infty,\infty)$. The rate $\Gamma_{2 \to 1}$ for the \emph{reverse} transition
from 
$\boldsymbol{A}^\star_2$ to $\boldsymbol{A}^\star_1$ is
similarly $\Gamma_{2\to 1} \sim \exp(-\mathcal{S}^\star_{2\to 1}/r)$.

The attraction distributions we are after will consist of
narrow (for small $r$) peaks
at $\boldsymbol{A}^\star_1$ and $\boldsymbol{A}^\star_1$. The weights $\omega_1$ and $\omega_2$ of these two peaks, which represent the probability for an agent to be within each peak, must then be such that forward and backward transitions balance:
\begin{align}
  \label{eq:dbFreidlinWentzell}
  \omega_1  \Gamma_{1 \to 2} &= \omega_2  \Gamma_{2 \to 1}\\
\frac{\omega_1}{\omega_2} & \propto \exp\left(\frac{\mathcal{S}^\star_{1\to 2} - \mathcal{S}^\star_{2\to 1}}{r}\right)
\end{align}
This expression shows that when the forward and backward minimal
actions are not equal, then one of the two peaks will
have an exponentially small weight as $r\to 0$. In practice this is true when the action difference inside the exponential in \eqref{eq:dbFreidlinWentzell} is large compared to $r$. If it is only of order $r$ or smaller, then we cannot say anything about the weights as we do not determine prefactor in \eqref{eq:dbFreidlinWentzell}, though we would expect them to be of order unity.

\subsubsection*{Finding the minimal action path numerically}

Following the method of Bunin \etal{}~\cite{bunin2012large}, we find
the minimal action by discretizing the path $[\boldsymbol{A}(t)]$,
evaluating the action as a function of this discretized path and then minimizing with respect to the (discretized) path. The path is discretized into 10 equally spaced timesteps between $t = 0$ and $t=10$; we found this choice of parameters to be a reasonable trade-off between the precision of our result and the complexity of minimizing the discretized action.

There are other methods for finding the minimal value of the action defined in
Eq.~\eqref{eq:action}, such as solving a Hamilton-Jacobi
equation~\cite{bouchetetal}, but we chose to use the path
discretization method because we found this to be more robust with respect to changes of model parameters. The discretization approach could also be improved further, using for example the geometric minimum action method~\cite{heymann2008pathways}, but we found that this was not necessary to achieve the desired precision. We tested this e.g.\ by benchmarking against closed-form results that can be obtained for $\alpha=1$~\cite{aloric2015emergence}.

\begin{figure}[t!!] \centering
  \includegraphics[scale=0.7]{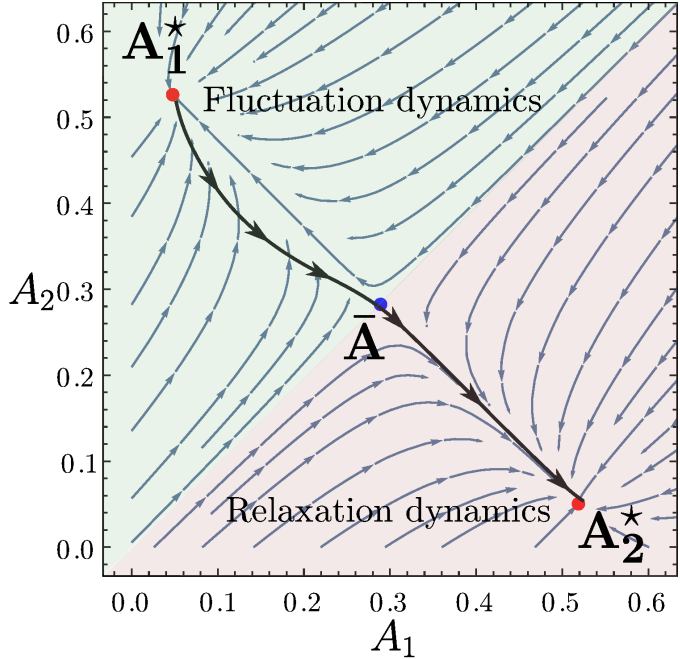}
  \caption{An example of a minimal action path, from fixed point $\boldsymbol{A}^\star_1$ to $\boldsymbol{A}^\star_2$. The path starts with a ``fluctuation" (or: activation) segment that ends at the saddle point $\bar{\boldsymbol{A}}$ 
    between the two fixed points. The remainder of the path is a ``relaxation'' segment that follows  the deterministic dynamics and incurs zero contribution to the action.}
  \label{fig:traj}
\end{figure}

The numerical path optimization can be simplified by restricting attention to the \emph{activation} part of the path. Generally, for a system with two stable fixed points
$\boldsymbol{A}^\star_1$ and $\boldsymbol{A}^\star_2$ and one saddle
point $\bar{\boldsymbol{A}}$ between them, the optimal path starting from $\boldsymbol{A}^\star_1$ 
will pass through the saddle point $\bar{\boldsymbol{A}}$ and  then relax to $\boldsymbol{A}^\star_2$ following the relaxation dynamics
$\dot{\boldsymbol{A}}(t) = \pmb{\mu}(\boldsymbol{A} (t))$, 
as sketched in Fig.~\ref{fig:traj}~\cite{fwtheory}. Eq.~\eqref{eq:action} shows that the
relaxation dynamics does not contribute to the total action as the integrand (the Lagrangian) vanishes identically along this section of the path. As a consequence,
the problem of finding a minimal action path
between $\boldsymbol{A}^\star_1$ and $\boldsymbol{A}^\star_2$ 
can be reduced to finding the minimal action path between $\boldsymbol{A}^\star_1$
and $\bar{\boldsymbol{A}}$, \ie\ from the initial fixed point to the saddle.
This restriction significantly improves the precision of the numerical path optimization.

With the above method, we can work out the action difference between any two fixed points of the single agent dynamics, as a function of the aggregates $\bar{p}^{(1)}$, $\bar{p}^{(2)}$; only the first of these is needed for symmetric steady states.  The values of $\bar{p}^{(1)}$ where the action difference between two single agent fixed points vanishes identify the points where the steady state attraction distribution of EWA learning can have more than one peak. Either side of these values, a single peak is dominant in the attraction distribution; which peak this is changes discontinuously at a zero action difference value of $\bar{p}^{(1)}$, see Fig.~\ref{fig:selfcon}.

\section{Conclusion}
\label{sec:conclusion}
In this paper we studied
a
minimal model of agents choosing between two double auction markets, which is a special case of a
large aggregative game. 
Previous work studying EWA (experience weighted attraction) learning in this system had found segregation, where a group of identical agents becomes heterogeneous by separating into sub-groups adopting different behaviours. We first asked the question of whether this phenomenon has an analogue in the Nash equilibria of the corresponding game, where -- in contrast to the EWA dynamics -- agents have full information about their expected payoffs.

In a game theoretical analysis (Sec.~\ref{sec:mf}) we addressed this question within a setup where there are two classes of agents that typically buy and sell, respectively. We showed
that two \emph{aggregate} quantities, namely, the fraction of agents from each class choosing the first market, are sufficient to assess whether a distribution of strategies, \ie\ market preferences, across the agents in each class is a Nash
equilibrium or not. This allowed us to classify the Nash equilibria, according to the type of strategies played by the agents (pure or mixed) and according to the distribution of strategies being  homogeneous (the entire class population
plays the same strategy) or heterogeneous (the population is
divided into subpopulations playing different strategies). The model
parameters for which each of these Nash equilibria exists
are summarized in Fig.~\ref{fig:phidiag}. A key conclusion is that there are regions of heterogeneous equilibria: these are the equilibrium analogues of dynamical segregation as observed previously.

This answer to our first question had to be qualified, however, because there is in general an infinity of strategy distributions consistent with a given pair of aggregate values. The Nash equilibrium analysis can therefore only identify equilibria as \emph{potentially}
heterogeneous but leaves open the nature of the actual strategy distribution, which could be homogeneous mixed, heterogeneous pure or heterogeneous mixed. We therefore asked a second question of whether EWA learning can resolve this ambiguity, by identifying which Nash equilibria can be reached dynamically.
We first argued that steady states of EWA learning should be Nash equilibria 
in the limit of perfect fictitious play
($\alpha \to 0$), long agent memory ($r \to 0$) and best response
($\beta \to \infty$).
(Sec.~\ref{sec:assum}).
Non-trivially, however, this joint limit can be taken in several ways, as shown in the phase
diagram in Fig.~\ref{fig:alphac}: depending on how
the point $(\alpha,1/\beta)=(0,0)$ is approached, a small number of different limiting steady states of EWA learning can result as sketched in Fig.~\ref{fig:diff_ne}.
These include a homogeneous mixed state, where all agents within a class randomize between markets in the same way, and a heterogeneous pure equilibrium, where agents separate into two groups, each choosing a market deterministically. Along with these standard types of Nash equilibria, however, we also found a \emph{heterogeneous mixed} steady state, where the agents do split into groups but not all groups play deterministically. In fact, at the boundary between the latter two types of steady states (denoted $\alpha=\alpha_{\rm c}'$ in our analysis) it is possible to generate equilibria where \emph{three} groups of agents appear within each class.

Technically what made our theoretical analysis of the heterogeneous steady states possible was the use of Freidlin-Wentzell theory, which is the tool of choice for studying the behavior of dynamical systems subject to weak noise, here arising from the limit $r \to 0$. We also compared the theoretical results to 
multi-agent simulations for $r>0$, finding good qualitative agreement.

While we focused our analysis on the study of the minimal model
of choice between double auction market presented in Sec.~\ref{sec:model}, our methods could be applied fruitfully also to the study 
of EWA learning in other types of
aggregative games such as the Cournot model~\cite{cournotappl}. It would be particularly interesting to see whether also here dynamical considerations single out particular Nash equilibria, including ones with the novel
heterogeneous mixed character that we found in our system.

At a technical level, future work could look more closely at the limit of large intensity of choice $\beta$ required to realize Nash equilibria as dynamical steady states. We approached this limit numerically, finding good agreement with theoretical predictions already for relatively modest $\beta$. An interesting challenge would be to take the full $\beta\to\infty$ limit in closed form within the analysis: preliminary work suggests that the large deviation analysis then becomes rather intricate, hence we leave this aspect for future work.

\bibliographystyle{plain} \bibliography{mfarticle}
\appendix
\section{Formula for the payoff}
\label{app:PayoffFormula}
To work out the average payoff of an ask (${\rm a}$) or bid (${\rm b}$) at market
$m$, we find first the probability for such an order to be
valid:
\begin{align}
  \mathcal{V}({\rm a},m) &= \mathbb{P}(\text{ask price $< \pi_m$}) = \frac{1}{\sqrt{2 \pi} \sigma} \int^{\pi_m}_{-\infty} \exp\left(- \frac{(x -\mua)^2}{2 \sigma^2}\right) {\rm d}x \\
  \mathcal{V} ({\rm b},m)&= \mathbb{P}(\text{bid price $> \pi_m$})= \frac{1}{\sqrt{2 \pi} \sigma} \int_{\pi_m}^{\infty} \exp\left(- \frac{(x - \mub)^2}{2 \sigma^2}\right) {\rm d}x
\end{align}
where the trading price $\pi_m$ is defined in equation \eqref{eq:tp}.

Once an order has been validated, it needs to be matched with that of a trader on the other side of the market. We denote the probability for this to happen for an order of type $\tau$ at market $m$ by $\mathcal{M}(\tau,m,f_m)$. This quantity depends on the ratio of the number of buyers and sellers in the market, 
$f_m = \frac{\text{\# buyers @ market m}}{\text{\# sellers @ market m}}$, as follows:
\begin{align}
  \mathcal{M}(\mathrm{a},m,f_m) &= \min \left( \frac{f_m \mathcal{V} ({\rm b},m)}{\mathcal{V} ({\rm a},m)},1\right) \label{eq:apm1}\\
  \mathcal{M}(\mathrm{b},m,f_m) &= \min \left( \frac{\mathcal{V} ({\rm a},m)}{f_m\mathcal{V} ({\rm b},m)},1\right)\label{eq:apm2}
\end{align}
where the first ratio in the minimum is that of the number of \emph{valid} buy and sell orders, always assuming large $N$ where fluctuations of these numbers can be neglected.

We call $\langle \mathcal{S}_{\tau,m}\rangle$
the average score of an order of type $\tau$, once it has been validated and successfully matched.
This is given by:
\begin{align}
  \langle \mathcal{S}_{{\rm a},m}\rangle &= \frac{1}{\mathcal{V}({\rm a},m)}\frac{1}{\sqrt{2 \pi} \sigma}
                                           \int^{\pi_m}_{-\infty} \left(\pi_m -x \right) \exp\left(- \frac{(x - \mua)^2}{2 \sigma^2}\right)  {\rm d}x \\
  \langle \mathcal{S}_{{\rm b},m}\rangle &= \frac{1}{\mathcal{V}({\rm b},m)}\frac{1}{\sqrt{2 \pi} \sigma}
                                           \int_{\pi_m}^{\infty}  \left(x - \pi_m \right) \exp\left(- \frac{(x - \mub^2)}{2 \sigma^2}\right) {\rm d}x
\end{align}
For later use we also define the average square of the score:
\begin{align}
  \langle \mathcal{S}^2_{{\rm a},m}\rangle &= \frac{1}{\mathcal{V}({\rm a},m)}\frac{1}{\sqrt{2 \pi} \sigma}
                                           \int^{\pi_m}_{-\infty} \left(\pi_m -x \right)^2 \exp\left(- \frac{(x - \mua)^2}{2 \sigma^2}\right)  {\rm d}x \\
  \langle \mathcal{S}^2_{{\rm b},m}\rangle &= \frac{1}{\mathcal{V}({\rm b},m)}\frac{1}{\sqrt{2 \pi} \sigma}
                                           \int_{\pi_m}^{\infty}  \left(x - \pi_m \right)^2 \exp\left(- \frac{(x - \mub)^2}{2 \sigma^2}\right) {\rm d}x
\end{align}
We can now compute the average payoff of an order of type $\tau$ at market $m$:
\begin{align}
  \mathcal{P}_{\tau,m}(f_m) = \mathcal{V}(\tau,m) \mathcal{M}(\tau,m,f_m)  \langle \mathcal{S}_{\tau,m}\rangle
\end{align}
Similarly, the average squared payoff that will appear in the second order moment of the Kramers-Moyal expansion in App.~\ref{app:kmexp} can be expressed as
\begin{align}
  \mathcal{Q}_{\tau,m}(f_m) &= \mathcal{V}(\tau,m) \mathcal{M}(\tau,m,f_m)  \langle \mathcal{S}^2_{\tau,m}\rangle\\
 \mathcal{Q}^{(c)}_{m}(f_m) &= \pb^{(c)}  \mathcal{Q}_{{\rm b},m}(f_m) + (1- \pb^{(c)}) \mathcal{Q}_{{\rm a},m}(f_m)
\end{align}
The second version here is averaged over the preference for buying and selling of an agent in class $c$.

\section{Phase diagram boundaries in
  Fig~\ref{fig:phidiag}}
\label{app:BoundNe}
In this section we indicate how to calculate phase boundaries in Fig.~\ref{fig:phidiag}, which shows the phase diagram for the case where the market bias and the probability to buy are symmetric ($\theta_1 = 1 - \theta_2$, $\pb \doteq \pb^{(1)} = 1 - \pb^{(2)} $).

At this boundary, a (symmetric) potentially heterogeneous Nash equilibrium (green triangle in Fig.~\ref{fig:eqpayoff}) turns smoothly into a homogeneous pure equilibrium (blue diamond and orange square in Fig.~\ref{fig:eqpayoff}) where the two classes of players choose different markets. One can therefore calculate the boundary by establishing the zone in the phase diagram where this homogeneous Nash equilibrium exists. For definiteness we consider the equilibrium $(\bar{p}^{(1)},\bar{p}^{(2)})=(1,0)$; the calculation for $(0,1)$ is completely analogous.

To get rid of the $\min$ in Eq.~(\ref{eq:apm1},~\ref{eq:apm2}) we focus in addition on the case where market 1
is saturated with sellers:
\begin{equation}
  \frac{f_1\mathcal{V}({\rm b},1)}{\mathcal{V} ({\rm a},1)} < 1
  \label{eq:appmsat}
\end{equation}
As a consequence the $\min$ term disappears from the market
conditions:
\begin{align}
  \mathcal{M}(\mathrm{b},1,f_m) &=\mathcal{M}(\mathrm{a},2,f_1) = 1 \\
  \mathcal{M}(\mathrm{a},1,f_m) &= \mathcal{M}(\mathrm{b},2,f_2) =  \frac{f_1 \mathcal{V} ({\rm b},1)}{\mathcal{V} ({\rm a},1)}
\label{eq:Match_probs}
\end{align}
Here the equality between $\mathcal{M}(\mathrm{a},1,f_1)$ and
$\mathcal{M}(\mathrm{b},2,f_2)$ comes from the symmetry of the
parameters. Because $(\bar{p}^{(1)},\bar{p}^{(2)})=(1,0)$, all agents from class 1  go to market 1 and so the buyer-to-seller ratios $f_m$ from \eqref{eq:2} are simple to express in terms of $\pb$:
\begin{align}
  f_1 = \frac{1}{f_2} = \frac{\pb}{1-\pb}
\end{align}

The payoffs at the two markets for traders from class 1 simplify accordingly:
\begin{align}
  \mathcal{P}^{(1)}_1(f_1
) &= \pb  \mathcal{V}(\mathrm{b},1) \langle S_{{\rm b,1}}\rangle + (1- \pb ) \mathcal{V}(\mathrm{a},1)\left[\frac{\pb}{1-\pb} \frac{\mathcal{V}(\mathrm{b},1)}{\mathcal{V}(\mathrm{a},1)}\right]  \langle S_{{\rm a,1}}\rangle \label{eq:apppo1}\\
  \mathcal{P}^{(1)}_2(f_1
) &= (1-\pb)  \mathcal{V}(\mathrm{a},2) \langle S_{{\rm a,2}}\rangle +  
\pb \mathcal{V}(\mathrm{b},2) \left[\frac{\pb}{1-\pb} 
\frac{\mathcal{V}(\mathrm{b},1)}{\mathcal{V}(\mathrm{a},1)}\right]
\langle S_{{\rm b,2}}\rangle \label{eq:apppo2}
\end{align}
The factors in brackets are the matching probabilities from \eqref{eq:Match_probs}, from which $\mathcal{V}(\mathrm{a},1)$ cancels in the first equation and
similarly (by symmetry) $\mathcal{V}(\mathrm{a},1) = \mathcal{V}(\mathrm{b},2)$ in the second.

\begin{figure}[h!]
  \centering
  \includegraphics[scale = 0.6]{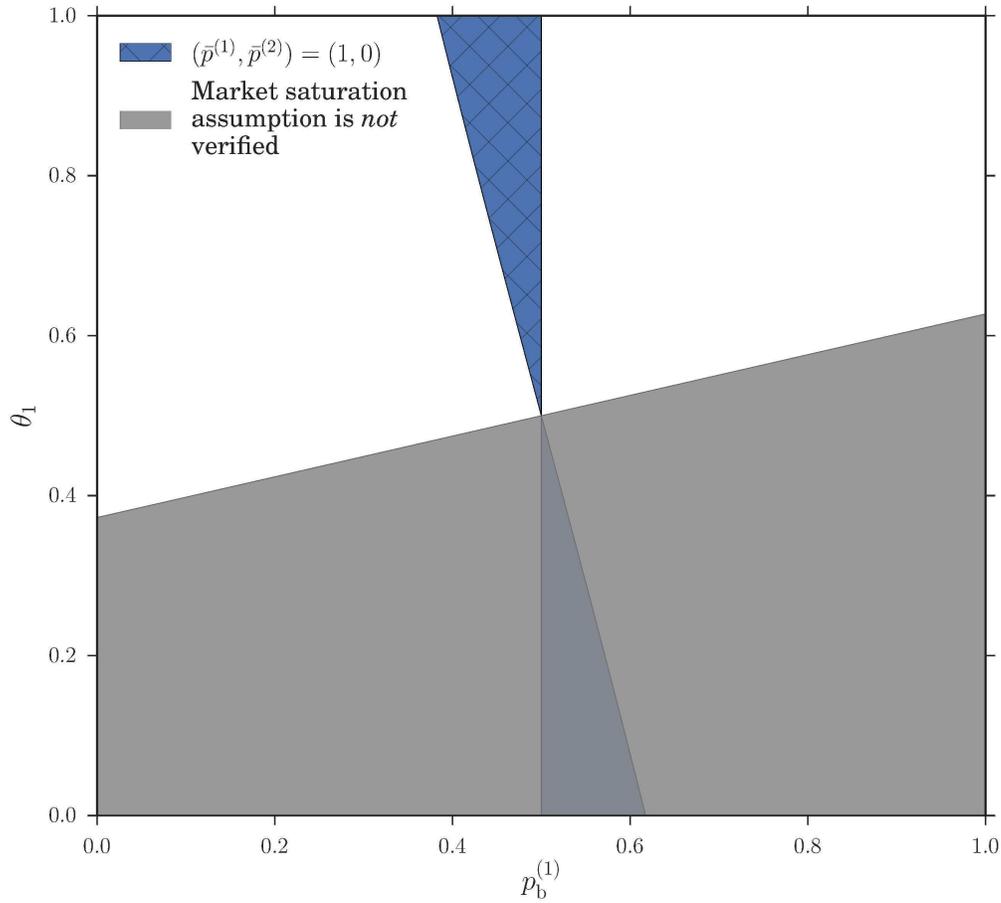}
  \caption{Analytic determination of boundaries for the zone where a homogeneous Nash equilibrium exists where players from the two classes choose different markets. Within the blue regions the payoff inequality \eqref{eq:secondordereq} is satisfied. The region shaded grey is ruled out by the assumption of market 1 being saturated with sellers.
}
  \label{fig:appBoundNe}
\end{figure}

Our assumed equilibrium $(\bar{p}^{(1)},\bar{p}^{(2)}) = (1,0)$ will be a Nash equilibrium if the payoff at market 1 is higher than at market 2 for 
players from class 1. (By symmetry, the payoff relation is then reversed for players in class 2.)
From the explicit payoff expressions above, this condition can be re-arranged into
\begin{align}
  \label{eq:secondordereq}
  0  &\le \pb^2 \left( -  \langle S_{{\rm a},2} \rangle  \mathcal{V}(\mathrm{a},2)-  \langle S_{{\rm b},1}\rangle  \mathcal{V}(\mathrm{b},1) -  \langle S_{{\rm a},1} \rangle \mathcal{V}(\mathrm{b},1) -  \langle S_{{\rm b},2} \rangle\mathcal{V}(\mathrm{a},2)\right) \notag\\
   &+ \pb \left(  \langle S_{{\rm b},1}\rangle  \mathcal{V}(\mathrm{b},1) + 2  \langle S_{{\rm a},2} \rangle  \mathcal{V}(\mathrm{a},2) +  \langle S_{{\rm a,1}}\rangle \mathcal{V}(\mathrm{b},1) \rangle\right) -   \langle S_{{\rm a},2} \rangle  \mathcal{V}(\mathrm{a},2)
\end{align}
For given $\theta_1$ all coefficients in this quadratic equation are known so the phase boundaries can be obtained directly as its roots. 
We plotted these roots in Fig.~\ref{fig:appBoundNe}; note that the boundaries are close to linear but not exactly so.
One has to check a posteriori that the assumption \eqref{eq:appmsat} of market 1 being saturated with sellers is valid, which rules out the bottom ``cone'' in the figure.

The remainder of the phase diagram in Fig.~\ref{fig:phidiag} is obtained by the analogous calculation under the assumption that market 1 is saturated with buyers rather than sellers, which yields the bottom ``cone'' in Fig.~\ref{fig:appBoundNe} and by finally repeating the overall reasoning for the Nash equilibrium $(\bar{p}^{(1)},\bar{p}^{(2)}) = (0,1)$.

\section{Kramers-Moyal expansion}
\label{app:kmexp}
Here we provide the coefficients of the Kramers-Moyal expansion for
traders with fixed buy-sell preference, given fictitious play
coefficient $\alpha$ and intensity of choice $\beta$.  The truncation of the Kramers-Moyal expansion at the
second order gives the Fokker-Planck equation for the time evolution of the attraction distributions:
\begin{align}
  \label{eq:app:km}
  \partial_t \mathbb{P}(\mathbf{A}^{(c)},t) &=
  - \sum_{1\le m \le 2  } \partial_{A^{(c)}_m}[\mu^{(c)}_m(\mathbf{A}^{(c)},\bar{p}^{(1)},\bar{p}^{(2)})\mathbb{P}(\mathbf{A}^{(c)},t)]\notag \\
  &+ \frac{r}{2} \sum_{1\le m,m' \le 2}\partial^2_{A^{(c)}_m A^{(c)}_{m'}} [\Sigma^{(c)}_{m m'}(\mathbf{A}^{(c)},\bar{p}^{(1)},\bar{p}^{(2)})\mathbb{P}(\mathbf{A}^{(c)},t)]
\end{align}
To lighten the notation we will in the following drop the superscript $(c)$ indicating the class of an agent
and also suppress the dependence on the aggregates $\bar{p}^{(1)},\bar{p}^{(2)}$, which are in general time-dependent via Eq.~\eqref{eq:detdynaggr}.

In the above expansion time has been rescaled as $t=rn$, where $n$ is the number of trading rounds. The time interval $\Delta t = r$ then features in the normalization of the drift and diffusion matrix, which are determined as the first and second order jump moments:
\begin{equation}
\boldsymbol{\mu} = \frac{1}{r} \langle \Delta \mathbf{A}\rangle, \qquad
r\mathbf{\Sigma} = \frac{1}{r} \langle \Delta \mathbf{A}\,
\Delta \mathbf{A}^{\rm T}\rangle
\end{equation}
where $\Delta \mathbf{A}=\mathbf{A}(n+1)-\mathbf{A}(n)$ is the change in the agent's attraction vector in one training round and the T superscript indicates vector transpose. Writing $\Delta\mathbf{A}$ explicitly from \eqref{eq:EWA_dynamics} then gives for the drift term:
\begin{align}
  \mu_1(\mathbf{A}) &= [\mathcal{P}_1(f_1) -  A_1] \sigma_\beta(A_1- A_2) -\alpha A_1 \sigma_\beta(A_2- A_1) \label{eq:appdr1}\\
  \mu_2(\mathbf{A}) &= [\mathcal{P}_2(f_2) -  A_2] \sigma_\beta(A_2- A_1) -\alpha A_2 \sigma_\beta(A_1- A_2) \label{eq:appdr2}
\end{align}
In the diffusion term $\Sigma_{ij}$ the second order
moments of the score distribution
also feature, as follows: 

\begin{align}
  \Sigma_{11} (\mathbf{A}) &= \left[\mathcal{Q}_1(f_1)  - 2A_1 \mathcal{P}_1(f_1) + {A_1}^2\right]  \sigma_\beta(A_1- A_2)+
                                                                   \alpha^2 {A_1}^2  \sigma_\beta(A_2- A_1)\\
  \Sigma_{22} (\mathbf{A}) &= \left[\mathcal{Q}_2(f_2)  - 2A_2 {\mathcal{P}_2}(f_2) + {A_2}^2\right]  \sigma_\beta(A_2- A_1)+ 
  \alpha^2 {A_2}^2  \sigma_\beta(A_1- A_2) \\
  \Sigma_{12} (\mathbf{A}) & = -\alpha \left[\mathcal{P}_1(f_1) A_2 \sigma_\beta(A_1- A_2) +
 \mathcal{P}_2(f_2) A_1\sigma_\beta(A_2- A_1) 
-A_1A_2\right]\\
  \Sigma_{21} (\mathbf{A}) &=  \Sigma_{12} (\mathbf{A})
\end{align}

\section{Fixed points of single agent dynamics}
\label{app:scalingbeta}

We show here generally that the single agent dynamics can have up to five fixed points, which can be determined from a single nonlinear equation. As before we drop the superscript $(c)$ for the agent class. The aggregates and hence the expected payoffs $\mathcal{P}_1$, $\mathcal{P}_2$ are fixed.

\begin{figure}[t!!]
  \centering
  \includegraphics[scale = 0.5]{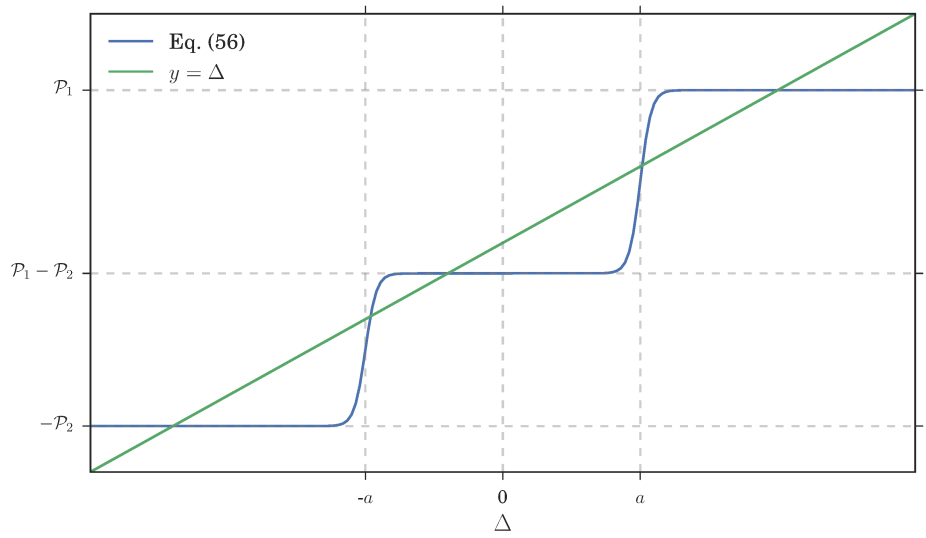}
  \caption{Sketch of the right hand side of the fixed point equation \eqref{eq:appendix1} for $\Delta$}
\label{fig:rhsapp}
\end{figure}

Fixed points are found from the condition that the drift (\ref{eq:appdr1},\ref{eq:appdr2})
must vanish: 
\begin{align}
  0 &= (\mathcal{P}_1 -  A_1) \sigma_\beta(A_1- A_2) -\alpha A_1 \sigma_\beta(A_2- A_1) \\
  0 &= (\mathcal{P}_2 -  A_2) \sigma_\beta(A_2- A_1) -\alpha A_2 \sigma_\beta(A_1- A_2)
\end{align}
Writing $\Delta=A_1-A_2$ and using $\sigma_\beta(A_2-A_1)=1-\sigma_\beta(\Delta)$, one can express $A_1$ and $A_2$ in terms of $\Delta$:
\begin{eqnarray}
A_1 &=& \frac{\mathcal{P}_1\sigma_\beta(\Delta)}{
\sigma_\beta(\Delta) + \alpha[1-\sigma_\beta(\Delta)]} \ =\  
\frac{\mathcal{P}_1}{
1+ \alpha \exp(-\beta\Delta)}
\\
A_2 &=& \frac{\mathcal{P}_2[1-\sigma_\beta(\Delta)]}{
1-\sigma_\beta(\Delta) + \alpha \sigma_\beta(\Delta)}
 \ =\  
\frac{\mathcal{P}_2}{
1+ \alpha \exp(\beta\Delta)}
\end{eqnarray}
Taking the difference gives a single equation for $\Delta$, which takes a suggestive form if we write $\alpha = \exp(-a\beta)$:
\begin{equation}
  \Delta = \frac{\mathcal{P}_1}{1 + \exp(-\beta (\Delta+a))} - \frac{\mathcal{P}_2}{1 + \exp(\beta (\Delta-a))}
  \label{eq:appendix1}
\end{equation}
The solutions of this equation, and hence the single agent fixed points, can be obtained graphically by intersecting a straight line (the l.h.s.\ 
of Eq.~\eqref{eq:appendix1}) with the function of $\Delta$ on the r.h.s. This function has a simple shape as it is the sum of two sigmoids, one increasing from zero to $\mathcal{P}_1$ around $\Delta=-a$ and the other increasing from $-\mathcal{P}_2$ to zero around $\Delta=a$. From the resulting shape, shown in Fig.~\ref{fig:rhsapp}, at most five intersections with the diagonal can occur.

We are most interested in the limit of large 
intensity of choice $\beta$, where the sigmoids become step functions. For small $\alpha$, \ie\ large $a$, the only solution is then $\Delta=\mathcal{P}_1-\mathcal{P}_2$. As $\alpha$ is increased and hence $a$ is decreased, the sigmoidal steps move closer to the origin, each creating an additional pair of solutions when $a$ equals the relevant payoff (see Fig.~\ref{fig:rhsapp}). For large $\beta$, one therefore has as transition from one to three (two stable, one unstable) fixed points at 
\begin{equation}
\alpha \sim \exp(-\max(\mathcal{P}_1,\mathcal{P}_2)\beta)
\end{equation}
and from three to five (three stable, two unstable) fixed points at
\begin{equation}
\alpha \sim \exp(-\min(\mathcal{P}_1,\mathcal{P}_2)\beta)
\end{equation}
At finite $\beta$ the fixed points are shifted away from $\Delta=\pm a$ and this would give corrections to $a$ of order $1/\beta$, which would in turn determine the prefactors of the above scalings.
Note that as $a$ decreases further, the two sigmoidal ramps will eventually overlap when $a$ is of order $1/\beta$, signalling a transition back to three (two stable) fixed points.

We show in Fig.~\ref{fig:alphac} that the scaling of the above $\alpha$-values, taken at equal payoffs $\mathcal{P}_1=\mathcal{P}_2$ as is relevant for Nash equilibria, also gives a good account of the variation with $\beta$ of $\alpha_c$ and $\alpha_c'$. 
This suggests that the $\alpha$-values where new fixed points appear, and where they contribute as peaks with weights of order unity to the steady state distribution, are relatively close, maybe only within a constant prefactor of each other.

\end{document}